\documentclass[12pt]{article}
\usepackage[utf8]{inputenc}

\usepackage{amsthm,amsmath,amsfonts,amssymb}
\usepackage[authoryear,round]{natbib}  
\usepackage[colorlinks=true, linkcolor=blue, citecolor=blue, urlcolor = blue, filecolor=blue]{hyperref}
\usepackage{graphicx}
\usepackage{xcolor}
\usepackage{subfiles}
\usepackage{multirow,multicol}
\usepackage{float}
\usepackage{algorithm} 
\usepackage{algpseudocode}
\usepackage[margin=1in]{geometry}
\usepackage{booktabs}
\usepackage{setspace}

\def\trans{^{\rm T}}
\DeclareMathOperator*{\argmax}{arg\,max}

\theoremstyle{plain}

\theoremstyle{remark}

\title{Integrative Learning of Quantum Dot Intensity Fluctuations under Excitation via Tailored Dynamic Mixture Modeling}
\author{
  Xin Yang\thanks{Department of Statistics, University of Connecticut, Storrs, CT, USA.}%
  \and
  Hawi Nyiera\thanks{Department of Chemistry, University of Connecticut, Storrs, CT, USA.}%
  \and
  Yonglei Sun\thanks{Institute of Materials Science, University of Connecticut, Storrs, CT, USA.}%
  \and
  Jing Zhao\footnotemark[2]%
  \and
  Kun Chen\footnotemark[1]
  \footnote{Corresponding author. Email: \texttt{kun.chen@uconn.edu}}
}
\date{\today}
\begin{document}
\maketitle

\begin{abstract}
Semiconductor nano-crystals, known as quantum dots (QDs),  have attracted significant attention for their unique fluorescence properties. Under continuous excitation, QDs emit photons with intricate intensity fluctuation: the intensity of photon emission fluctuates during the excitation, and such a fluctuation pattern can vary across different QDs even under the same experimental conditions. What adding to the complication is that the processed intensity series are non-Gaussian and truncated due to necessary thresholding and normalization. Conventional normality-based single-dot analysis fall short of addressing these complexities. In collaboration with chemists, we develop an integrative learning approach to simultaneously analyzing intensity series from multiple QDs. Motivated by the unique data structure and the hypothesized behaviors of the QDs, our approach leverages the celebrated hidden Markov model as its structural backbone to characterize individual dot intensity fluctuations, while assuming that, in each state the normalized intensity follows a 0/1 inflated Beta distribution, the state/emission distributions are shared across the QDs, and the state transition dynamics can vary among a few QD clusters. This framework allows for a precise, collective characterization of intensity fluctuation patterns and have the potential to transform current practice in chemistry. Applying our method to experimental data from 128 QDs, we reveal three shared intensity states and capture several distinct intensity transition patterns, underscoring the effectiveness of our approach in providing deeper insights into QD behaviors and their design and application potential.
\end{abstract}

\vspace{1em}
\noindent \textbf{Keywords:} Clustering; EM algorithm; Hidden Markov model; Integrative analysis; Zero-one inflation

\doublespacing

\section{Introduction}
In the fields of chemistry and material science, semiconductor Nano-crystals, commonly referred to as colloidal quantum dots (QDs), have gained significant attention in recent years. One captivating characteristic of the QDs is their ability to emit photons under photoexcitation, a phenomenon known as fluorescence. The applications of fluorescence property, particularly within the realm of optics and optoelectronics, have spurred extensive research and development efforts. For example, \citet{Bright2018_NaturePhoton} developed a process to construct more efficient quantum LEDs, and \citet{Bruns2017_NatureBE} used QDs as photostable probes to track and visualize biological tissues.

The importance of the fluorescence property in applications stems from its abilities to elucidate the intrinsic properties and micro-environment of QDs, particularly at the single QD level. Beyond the aforementioned applications, a more comprehensive understanding of the fluorescence phenomenon would pave the way for designing QDs with specific purposes and providing essential guidelines for the synthesis of novel materials. This urgent need amplifies our focus on a comprehensive and rigorous statistical analysis of the fluorescence of the QDs. 

\begin{figure}[tp]
  \begin{center}
  \includegraphics[width=\linewidth]{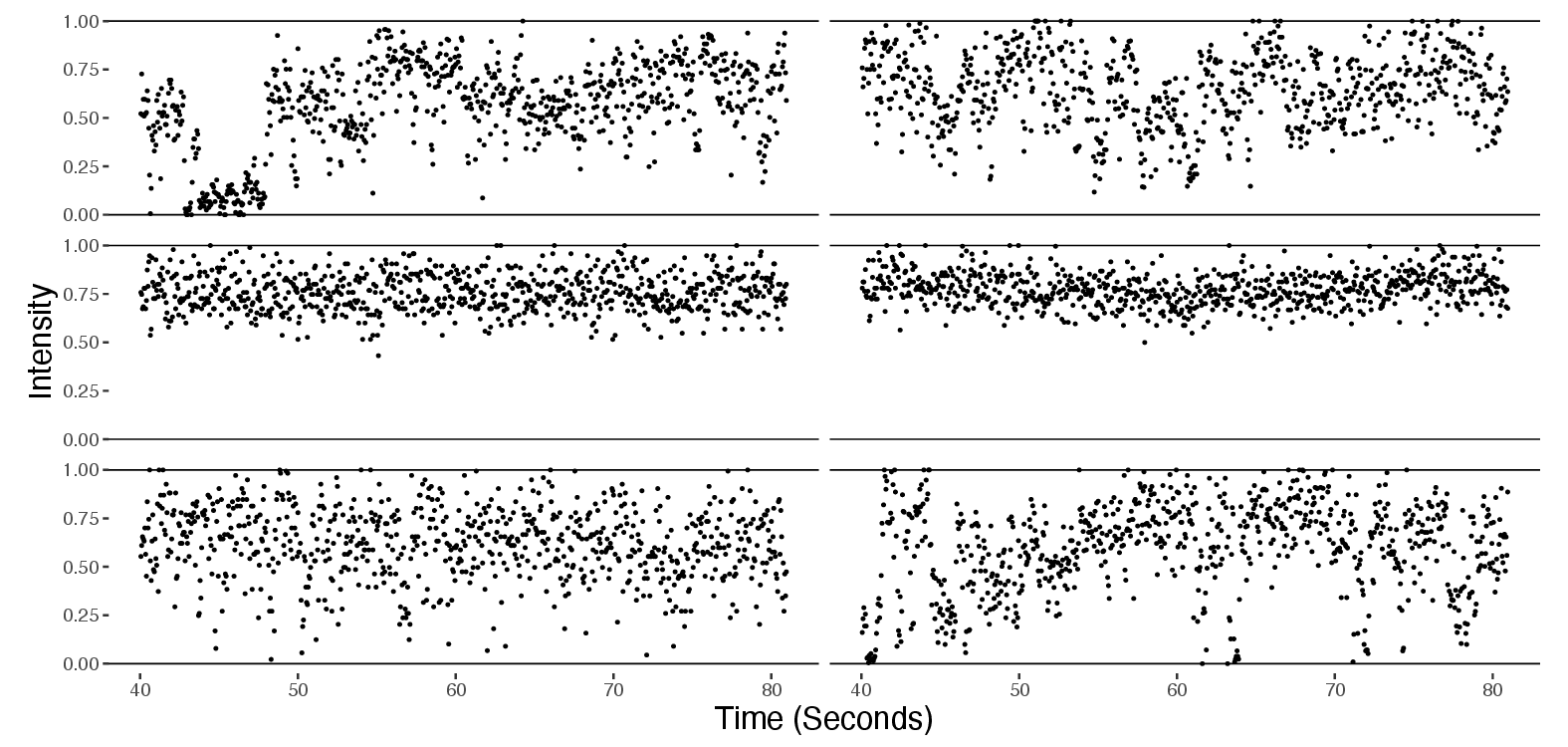}
  \end{center}
  \caption{Quantum dot data: Intensity fluctuation patterns in from some selected quantum dots.}
  \label{fig:Fluctuationtypes}
\end{figure}

An intriguing phenomenon of the fluorescence of QD is the so-called ``Intensity Intermittency'' or ``Intensity Fluctuation'', which refers to the observation that the intensity of photon emission fluctuates during the excitation process, and such fluctuation patterns can vary across different QDs even under the same experimental conditions. Some dynamics or patterns of intensity fluctuations \citep{Nirmal1996_Nature} have been observed across various QD types and have been well recognized in chemistry literature. For example, ``Intensity Blinking'' describes the pattern in which the intensity shows telegraph-like switching between high and low intensity states. ``Intensity Flickering'' has also been reported, in which the intensity exhibits frequent transitions across multiple intensity states. In reality, however, the intensity fluctuation patterns of single QDs are more complex and are not often described with scientific rigor. It is hypothesized that other patterns may exist.  Notably, in our data, as depicted in Figure \ref{fig:Fluctuationtypes}, while some QDs exhibit clear behaviors of either ``Blinking'' or ``Flickering'', others may show characteristics of neither of these two known fluctuation types.

In chemistry literature, researchers often conduct single-dot analysis of raw intensity series data, with, e.g., Gaussian hidden Markov models (HMM) \citep{McKinney2006_Biophy}. As the overall intensity level varies across QDs, the resulting high, medium, and low intensity states from HMMs could also vary across QDs. Their selection, labeling, and correspondence can then be arbitrary and often rely on the experiences of the researcher. 

In more recent studies, some advanced analytical methods have been developed, encompassing probability distribution analysis \citep{Efros1997_PRL, Kuno2000_JCP}, burst variance analysis \citep{Frantsuzov2008_NatureP}, FRET two-kernel density estimator \citep{Evangelos2010_MinEnzymology}, and fluorescence correlation spectroscopy \citep{Magde1972_PRL, Michsch2018_NanoL}. While each of these methods focuses on elucidating the fluctuation patterns of individual QDs, they tend to overlook the aspect of similarity across different QDs. As such, they fall short in meeting the many analytical challenges and may fail to capture any novel yet rare fluctuation patterns among QDs. 
\begin{figure}[h]
\begin{center}
\includegraphics[width=\linewidth]{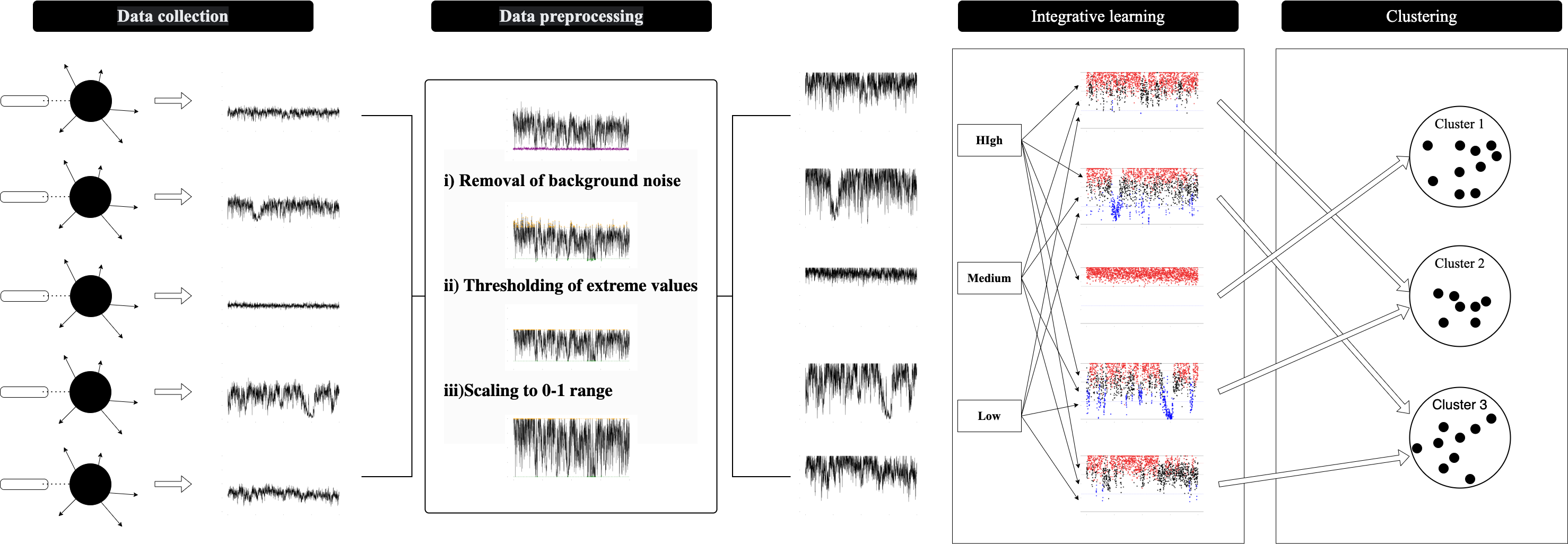}
\end{center}
\caption{Proposed pipeline of quantum dot analysis from data collection, data processing, to integrative learning with dynamic mixture model.}
\label{fig:Diagram1}
\end{figure}

Closely collaborating with scientists in chemistry, we innovate an integrative learning pipeline (as shown in Figure \ref{fig:Diagram1}) to simultaneously analyze standardized and robustified intensity series of multiple QDs. To achieve this, we first develop a data pre-processing procedure based on the nature of the experiment and limitations of the equipment, by eliminating background noise from the intensity values and thresholding and normalizing raw intensity values to remove the inherent variations between different QDs. After rescaling, the intensity values in each series then become bounded between 0 and 1 and can also be exactly equal to either 0 or 1. These processed intensities are then analyzed using our proposed integrative learning approach, in which state identification, transition pattern recognition, and QD clustering are performed simultaneously.

Our analytic approach still inherits the celebrated HMM as the skeleton, that is, for each dot, we assume the intensity fluctuations are governed by a set of unobserved hidden states that operate as a Markov chain. To enable integrative learning, we further assume that (1) under each state, the standardized intensity follows a 0/1 inflated Beta distribution based on empirical evidence, (2) the hidden state distributions are shared among all the QDs, and (3) the patterns of transitions can vary across QDs thus giving rise to a mixture HMM (MHMM). These features lead to a precise and collective characterization of the intensity fluctuation patterns and facilitate an objective clustering of the QDs. 

The remainder of the paper is organized as follows. In Section \ref{sec:2}, we describe the data and the problem setup for studying the intensity fluctuations. In Section \ref{sec:3}, after an overview of existing methods, we propose a dynamic mixture model with inflated Beta distribution, develop an efficient computational algorithm for parameter estimation, and discuss issues related to practical implementation, model selection, and inference. 
Extensive simulation studies are presented in Section \ref{sec:4}. In Section \ref{sec:5}, we thoroughly analyze the QD dataset and discuss the results and implications. In Section \ref{sec:6}, we provide a few concluding remarks and discuss future research directions.

\section{Data Description and Pre-Processing Pipeline}
\label{sec:2}
The data were compiled from 128 quantum dot samples. For each QD, we obtained 2000 equally-spaced intensity measurements of photon emission over a 100-second time frame (under continuous excitation). In the following, we detail the data acquisition and processing methodology. 

\subsection{Data Acquisition}
The QD samples, specifically $\mbox{CsPbBr}_3$ Perovskite Nanocrystals, were synthesized following the method from \citet{Protesescu2015_NanoL}. These samples were then diluted in a 3\% w/v polystyrene in toluene solution, deposited onto coverslips through spin casting, and examined under a Nikon Eclipse Ti-u microscope. Excitation of QDs was achieved using a 405 nm pulsed diode laser, with the resulting photo-luminescence filtered through a 510$\pm$40 nm band-pass filter. A time-correlated single photon counting module recorded the photo-luminescence in time-tagged time-resolved (TTTR) mode, capturing photon arrival time. Intensity was then calculated as the number of photons detected over a given time period divided by the length of that period. 

\subsection{Pre-Processing}
The raw intensity data that we obtained present several aspects of heterogeneity and irregularity, including the varying magnitudes of the intensity values between the QDs and the presence of outlying values.  We closely collaborated with chemists to design a data standardization procedure as follows.

\begin{enumerate}
\item Removal of background noise: For each QD, the average intensity of the background is estimated and subtracted from the raw intensity values.
\item Thresholding of extreme values: For each QD, the intensity threshold is set as the average of its 90th percentile and its maximum background-free intensity. Intensity values below zero are reset to zero, while those above the threshold are capped at the threshold value.
\item Scaling: For each QD, the intensity values are scaled to the 0-1 range by dividing its maximum value.
\end{enumerate}

We stress that the data preprocessing procedure described above is based on the nature of the chemical experiments and the limitations of the experimental equipment, and it is justified from the chemistry perspectives. This procedure yields a standardized dataset where each intensity series consists of 2000 equally-spaced intensity values for a 100-second time frame and with all the values ranging from 0 to 1 (inclusive). This makes a direct comparison across the QDs meaningful and thus enables their integrative learning. 

Table \ref{table:summaryIntensity} presents some summary statistics for the standardized intensity measurements. The results indicate that the mean intensity values of the QDs (averaged over time for each QD) are relatively concentrated, with a median of 0.694 and an interquartile range of 0.613 to 0.774. Additionally, the zero rate and the one rate remain low, with a maximum of 0.038, suggesting very limited occurrences of extreme values in the intensity values. These confirm that the standardized intensity values are comparable across the QDs. Furthermore, recall that Figure \ref{fig:Fluctuationtypes} shows that QDs may exhibit distinctive behaviors of intensity fluctuation over time. We therefore aim to perform an integrative learning of all QDs to reveal the unique yet interrelated intensity fluctuation mechanisms and potential clusters of the QDs. 

\begin{table}[t]
\centering
\caption{Quantum dot data: Summary statistics of standardized intensity series.}
\label{table:summaryIntensity}
\begin{tabular}{ c c  c c  c  c } 
 \hline
   & Min  & $25^{th}$ Percentile & Median & $75^{th}$ Percentile & Max \\ [0.5ex] 
 Mean Intensity & 0.480 & 0.613 & 0.694 & 0.774 & 0.833 \\ 
 Zero Rate & 0.000 & 0.000 & 0.000 & 0.000 & 0.038 \\
 One Rate  & 0.002 & 0.009 & 0.011 & 0.014 & 0.021 \\ 
 \hline
\end{tabular}
\end{table}

\section{Integrative Dynamic Mixture Model of Intensity Fluctuations}
\label{sec:3}
\subsection{Overview}

In the chemistry literature, researchers often analyze the dynamics and intensity fluctuations of each single QD. For example, \citet{McKinney2006_Biophy} introduced an HMM approach, assuming that different trajectories shared a homogeneous transition pattern. Consequently, their approach involved fitting an HMM model to each (representative) trajectory and calculating the average transition matrices. \citet{Pirchi2016_JPCB} utilized an $\mbox{H}^2\mbox{MM}$ framework, an enhanced version of HMM. The novelty of their approach lied in the consideration of hidden state transitions and non-transition times; two sets of hidden states were assigned -- one for transitions and another for non-transitions. 

These HMM based single-dot approaches allow for the classification of intensity levels into discrete states, such as high, medium, and low. However, challenges arise because intensity levels vary across QDs, leading to subjective labeling and reliance on researcher expertise for state correspondence. While HMMs provide valuable insights into individual QDs, they lack the capacity to analyze patterns across multiple QDs simultaneously, missing opportunities to identify shared behaviors or rare but novel anomalies.
 
To address data heterogeneity, finite mixture models \citep{Dempster1977,YaoXiang2024} have been widely adopted in various fields, including medicine \citep{Schlattmann1993}, public health \citep{Fahey2011_JECH}, genetics \citep{ChenMishra2017JASA}, psychology \citep{Steinley2011_PsycM}, and economics \citep{Deb2011_JHE, Deb1997_JAE}. These models decompose data into a weighted sum of distributions, each representing a cluster, thereby capturing group-level patterns.

The integration of HMMs and mixture models, particularly in the form of Mixture Hidden Markov Models (MHMMs), has shown promise in addressing temporal and cluster-level complexities. MHMMs extend HMMs by allowing the transition dynamics to vary across clusters, enabling simultaneous state identification and clustering. Historically, MHMMs have been mainly applied to Gaussian data, such as in neuroscience to classify EEG signals \citep{Wang2018_IEEE} or in economics to study market volatility \citep{Dias2010_Book}. However, Gaussian assumptions are often unsuitable for datasets like QD intensities, where distributions exhibit bounded behavior and inflation at extreme values (e.g., 0 and 1).

\subsection{Mixture HMM with 0/1 Inflated Beta}\label{sec:3:sub:2}
We consider a set of $N$ QD samples, where each QD is associated with intensity values at $T$ different time points. Let $x_{i,t}$ denote the intensity of $i^{th}$ QD at time $t$, where $i\in \{1,\ldots,N\}$ and $t\in \{1,\ldots,T\}$. Correspondingly, the vector $\boldsymbol{x}_{i} = (x_{i,1},\dots,x_{i,T})\trans$ denotes the intensity series of dot $i$ and the matrix $\boldsymbol{X} = (\boldsymbol{x}_1,\ldots,\boldsymbol{x}_N)\trans$ consists of all the observed QD intensity series.

Our approach inherits the HMM as the skeleton, that is, for each dot, we assume the intensity fluctuations are governed by a set of unobserved hidden states that operate as a Markov chain. 
With $M$ states, the state of the intensity $x_{i,t}$ can be represented as a vector $\boldsymbol{s}_{i,t} = (s_{i,t,1},\ldots,s_{i,t,M})\trans$, where $s_{i,t,h}$, $h=1,\ldots, M$, are indicator variables and $s_{i,t,h} = 1$ only if $x_{i,t}$ is at state $h$. 

We use $\boldsymbol{s}_i= (\boldsymbol{s}_{i,1},\ldots,\boldsymbol{s}_{i,T})\trans$ to represent the state series of $\boldsymbol{x}_i$ and use $\boldsymbol{S} = (\boldsymbol{s}_1,\ldots,\boldsymbol{s}_N)\trans$ to collect state series for all the intensity series. 

With the above general HMM setup, we impose several tailored structures for the integrative analysis of the QDs. First, we assume that under each state, the standardized intensity follows a 0/1 inflated Beta distribution. That is, the probability density function of the intensity under state $h$ is written as
\begin{equation}
\begin{aligned}
& f(x_{i,t} \mid s_{i,t,h} = 1, a_{i,h}, b_{i,h},\epsilon_{i,h}^{(0)}, \epsilon_{i,h}^{(1)}) \\
= & (\epsilon_{i,h}^{(0)})^{I(x_{i,t}=0)} [(1- \epsilon_{i,h}^{(0)} - \epsilon_{i,h}^{(1)}) f(x_{i,t} \mid a_{i,h},b_{i,h})]^ {I(0 < x_{i,t} < 1)}  (\epsilon_{i,h}^{(1)})^{I(x_{i,t} = 1)},
\end{aligned}\nonumber
\end{equation}
where $\epsilon_{i,h}^{(0)} = P(x_{it,}=0 \mid s_{i,t,h} = 1)$ and $\epsilon_{i,h}^{(1)} = P(x_{i,t}=1 \mid s_{i,t,h} = 1)$ are the inflated probabilities, and $f(x_{i,t}\mid a_{i,h},b_{i,h})$ is the density function of the Beta distribution, $\mbox{Beta}(a_{i,h},b_{i,h})$. 

Based on the chemical process, it is expected that each QD only has a few intensity states, in particular, three states representing relatively low, median, and high intensities. Since all data are standardized, we further assume that the possible intensity states are shared across all the QDs, that is, the parameters of the inflated Beta distributions no longer depend on $i$, so that the probability density function under state $h$, $h=1,\ldots,M$, can be simplified as
\begin{equation}
\begin{aligned}
& f(x_{i,t} \mid s_{i,t,h} = 1, a_{h}, b_{h},\epsilon_{h}^{(0)}, \epsilon_{h}^{(1)}) \\
= & (\epsilon_{h}^{(0)})^{I(x_{i,t}=0)} [(1- \epsilon_{h}^{(0)} - \epsilon_{h}^{(1)}) f(x_{i,t} \mid a_{h},b_{h})]^ {I(0 < x_{i,t} < 1)}  (\epsilon_{h}^{(1)})^{I(x_{i,t} = 1)}, 
\end{aligned}
\label{eq:INFLATEDBETA1}
\end{equation}
This effectively enables integrative learning, as now the potential intensity states are to be collectively identified from all the QDs. 

It is now in order to capture the heterogeneity in the intensity fluctuation across the QDs.  We achieve this by considering the transition patterns in the hidden Markov process and assuming that the state transition probabilities can vary across QDs. This gives rise to a mixture HMM (MHMM), with which all QDs can be clustered into $K$ clusters/groups, each characterized by a unique state transition or fluctuation pattern. Specifically, let the cluster number of each dot $i$ be represented by a vector $\boldsymbol{z}_i = (z_{i,1},\ldots,z_{i,K})\trans$, where $z_{i,k} = 1$ if dot $i$ belongs to group $k$ and otherwise $z_{i,k} = 0$, for $k=1,\ldots, K$, $i \in \{1,\ldots,N\}$, and let $\boldsymbol{Z} = (\boldsymbol{z}_1,\ldots,\boldsymbol{z}_N)\trans$.
Let $\delta_k = P(z_{i,k} = 1)$, $k=1,\ldots, K$, with $\sum_{k=1}^K\delta_k = 1$. For each cluster, we use $\boldsymbol{\Pi}_k$ to denote the transition matrix, where its element $(\boldsymbol{\Pi}_k)_{p,q}$ gives the probability of switching to state $q$ from state $p$, i.e., $ (\boldsymbol{\Pi}_k)_{p,q} = P(s_{i,t,q}=1 \mid s_{i,t-1,p}=1)$. We then use $\boldsymbol{\Pi}$ to denote the collection of all the transition matrices $\boldsymbol{\Pi}_k, k \in \{1,\ldots, K\}$.

\begin{figure}[t]
\begin{center}
\includegraphics[width=\linewidth]{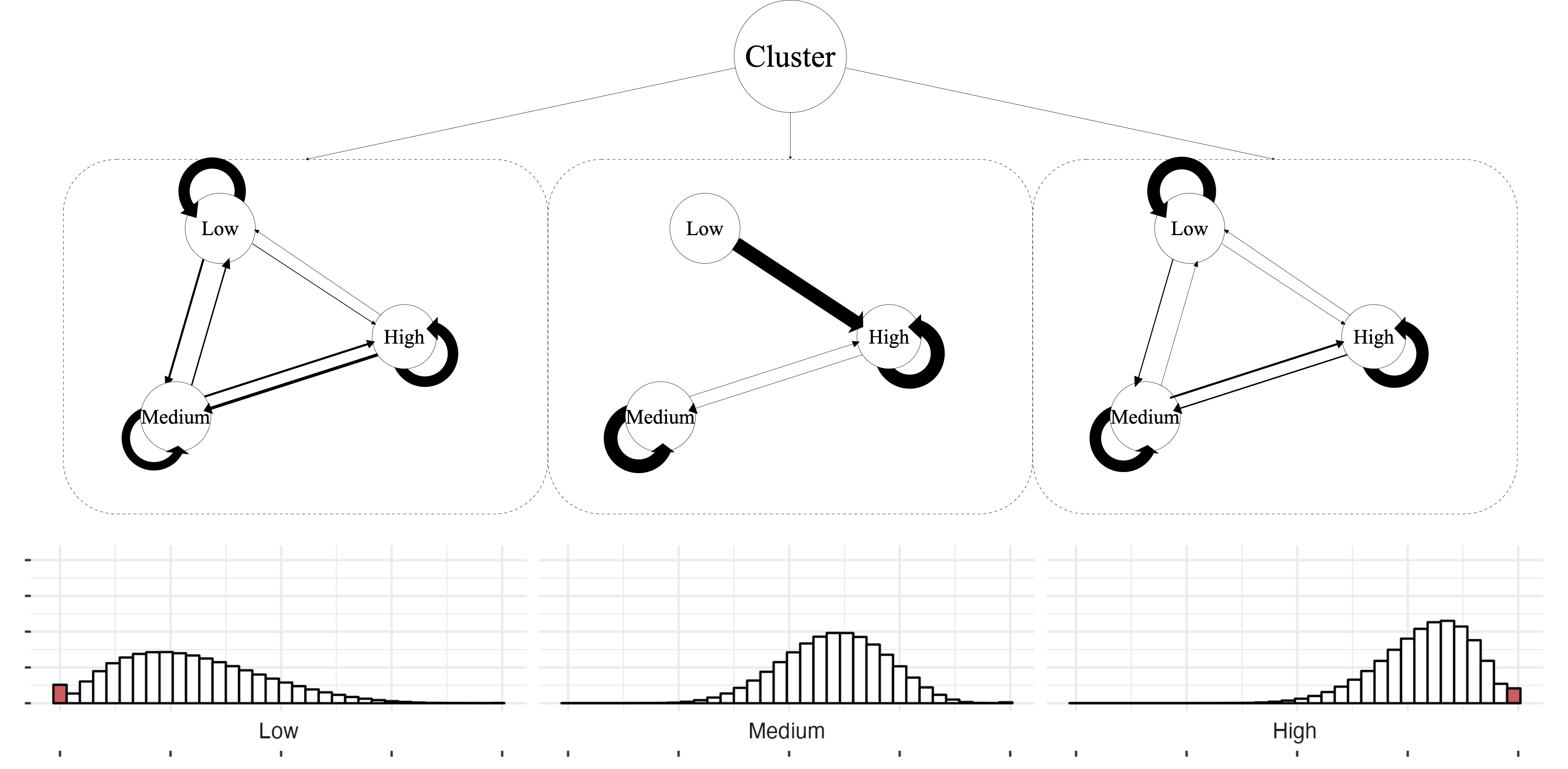}
\end{center}
\caption{ Schematic illustration of the integrative mixture hidden Markov model with 0/1 inflated Beta (MHMM-$\beta$).}
\label{fig:Diagram2}
\end{figure}

We term the proposed method as the \textit{integrative mixture hidden Markov model with 0/1 inflated Beta}, denoted as MHMM-$\beta$. The tailored model structures imposed above lead to a precise and collective characterization of the intensity fluctuation patterns and facilitate an objective clustering of the QDs. To summarize and visualize, Figure \ref{fig:Diagram2} shows a conceptual diagram of the proposed MHMM-$\beta$ model. The clusters share the same set of hidden states ($M=3$) and are distinguished by different transition patterns ($K=3$). The shared states, i.e., the three inflated Beta distributions, are shown by the three histograms at the bottom of the figure. The potential transitions are represented by the arrows, where the thickness indicates the magnitude of the corresponding probability. A QD in Cluster 1 tends to stay in its current state and rarely transits to other states, whereas a QD in Cluster 2 may always transit to the high state from the low state (essentially it means that the QD does not spend any time in the low state). 

\subsection{Likelihood Derivation and Estimation Criterion}

Here we derive the data likelihood function and discuss the maximum likelihood estimation of the parameters. To proceed, we use $\pi_{k,\boldsymbol{s}_{i,1}}$ to denote the probability of the initial state when the $i$th QD is in the $k$th cluster, such that $\pi_{k,h} = Pr(s_{i,1,h} = 1 \mid z_{i,k} = 1)$ and $\sum_{h=1}^M \pi_{k,h} = 1$. Let $\boldsymbol{\pi}_k=(\pi_{k,1},\ldots,\pi_{k,M})\trans$, and 
$\boldsymbol{\pi} = (\boldsymbol{\pi}_1,\ldots,\boldsymbol{\pi}_K)\trans$. Let $\boldsymbol{\Theta}$ denote the set of all the unknown parameters, including the parameters for the state distributions $(\epsilon_h^{(0)},\epsilon_h^{(1)},a_h,b_h)_{h = 1,\ldots,M}$, the parameters in the transition matrices $(\boldsymbol{\Pi}_k)_{k = 1,\ldots,K}$, and the parameters for the initial states $\boldsymbol{\pi}$.

Two key properties of the Markov chain are utilized. First, the sequence $(\boldsymbol{s}_{i,1},\ldots,\boldsymbol{s}_{i, T})$ forms a Markov chain if $\boldsymbol{s}_{i,t+1} \perp (\boldsymbol{s}_{i,1}, \ldots, \boldsymbol{s}_{i,t-1}) \mid \boldsymbol{s}_{i,t}$, meaning that the future is conditionally independent of the past given the present state. 
Second, the condition $x_{i,t} \perp (\boldsymbol{s}_{i,1},\ldots,\boldsymbol{s}_{i,t-1},x_{i,1},\ldots,x_{i,t-1}) \mid \boldsymbol{s}_{i,t}$ holds, meaning that the present observation depends only on the present state. With these properties and the inflated Beta distribution defined in \eqref{eq:INFLATEDBETA1}, the complete-data likelihood of each dot $i$, given the intensity series $\boldsymbol{x}_i$, the state series $\boldsymbol{s}_i$, and the cluster information $\boldsymbol{z}_i$, is given by
\begin{equation}
L(\boldsymbol{\Theta} \mid \boldsymbol{x}_i,\boldsymbol{s}_i,\boldsymbol{z}_i) \\
 = \prod_{k=1}^K \left\{ 
 \delta_k 
 \pi_{k,\boldsymbol{s}_{i,1}} f(x_{i,1} \mid \boldsymbol{s}_{i,1}) \prod_{t=2}^T (\boldsymbol{\Pi}_k)_{\boldsymbol{s}_{i,t-1},\boldsymbol{s}_{i,t}} f(x_{i,t} \mid \boldsymbol{s}_{i,t}) \right\}^{I(z_{i,k}=1)},
\label{fl:completeLikelihood}
\end{equation}
where $f(x_{i,t} \mid \boldsymbol{s}_{i,t})$ is the density function of the state, i.e., $f(x_{i,t} \mid s_{i,t,h} = 1) = f(x_{i,t} \mid s_{i,t,h} = 1, a_{h}, b_{h},\epsilon_{h}^{(0)}, \epsilon_{h}^{(1)})$ as defined in \eqref{eq:INFLATEDBETA1}. Then the complete-data likelihood function for all dots can be written as $L(\boldsymbol{\Theta} \mid \boldsymbol{X},\boldsymbol{S},\boldsymbol{Z})
 = \prod_{i=1}^N L(\boldsymbol{\Theta} \mid 
\boldsymbol{x}_i,\boldsymbol{s}_i,\boldsymbol{z}_i)$.

Both $\boldsymbol{s}_i$ and $\boldsymbol{z}_i$ are unobserved. From the complete-data likelihood function \eqref{fl:completeLikelihood}, the unobserved parts can be integrated out, leading to the observed data likelihood,
\begin{equation}
L(\boldsymbol{\Theta} \mid \boldsymbol{x}_i  )  = \sum_{k = 1}^K \left[
\delta_k 
\sum_{\boldsymbol{S}_{i} \in \mathcal{S}_{i}}\left\{ \pi_{k,\boldsymbol{s}_{i,1}} f(x_{i,1} \mid \boldsymbol{s}_{i,1}) \prod_{t=2}^T (\boldsymbol{\Pi}_k)_{\boldsymbol{s}_{i,t-1},\boldsymbol{s}_{i,t}} f(x_{i,t} \mid \boldsymbol{s}_{i,t}) \right\} \right],
\label{fl:observedlikelihood}
\end{equation}
where $\mathcal{S}_i$ stands for all the possible state sequences for dot $i$. As a key part to calculate the observed data log-likelihood, the summation of all possible sequences of hidden states is calculated using the forward probabilities \citep{Rabiner1989_IEEEproc}. More details can be found in Section \ref{sec:comp-ms}.

The likelihood function for all dots with the observed data is given by $L(\boldsymbol{\Theta} \mid \boldsymbol{X}) = \prod_{i=1}^N L(\boldsymbol{\Theta} \mid \boldsymbol{x}_i )$. We then propose to conduct the maximum likelihood estimation: 
\begin{equation}
  \widehat{\boldsymbol{\Theta}} = \argmax_{\boldsymbol{\Theta}} \log( L(\boldsymbol{\Theta} \mid \boldsymbol{X}))= \argmax_{\boldsymbol{\Theta}} l(\boldsymbol{\Theta} \mid \boldsymbol{X}).
  \label{fl:estimator}
\end{equation}
where $l(\boldsymbol{\Theta} \mid \boldsymbol{X}) = \sum_{i=1}^N \log L(\boldsymbol{\Theta} \mid \boldsymbol{x}_i  )$ is the log-likelihood function. 

\subsection{Computational Algorithm, Model Selection, \& Inference}\label{sec:comp-ms}

\subsubsection{Computational Algorithm}
We derive a generalized EM algorithm for optimization, which executes an E-step and an M-step iteratively until convergence. 

The procedures of the E-Step are shown as follows. First, we utilize the complete-data log-likelihood function in  \eqref{fl:completeLikelihood} and the multiple expectation structure to define the conditional expectation function of the complete-data log-likelihood as follows,
\begin{equation}
\begin{split}
Q(\boldsymbol{\Theta} \mid \boldsymbol{\Theta}^{(j)}) & =  E_{(\boldsymbol{S},\boldsymbol{Z}) \sim p(\cdot,\cdot \mid \boldsymbol{X},\boldsymbol{\Theta}^{(j)}) }[l(\boldsymbol{\Theta} \mid \boldsymbol{X},\boldsymbol{S},\boldsymbol{Z}) \mid \boldsymbol{X},\boldsymbol{\Theta}^{(j)} ] \\
& = E_{\boldsymbol{S}\sim p(\cdot \mid \boldsymbol{X},\boldsymbol{Z},\boldsymbol{\Theta}^{(j)}) }\{E_{\boldsymbol{Z} \sim p(\cdot \mid \boldsymbol{X},\boldsymbol{\Theta}^{(j)})}[ l(\boldsymbol{\Theta} \mid \boldsymbol{X},\boldsymbol{S},\boldsymbol{Z})  ]\},
\end{split}
\label{fl:Estep}
\end{equation}
where $\boldsymbol{\Theta}^{(j)}$ stands for the estimated parameters from $j$th iteration. In \eqref{fl:Estep}, the multiple integration part is simplified into a double expectation structure. As a prerequisite for the E-step, we adopt the idea of dynamic programming \citep{Rabiner1989_IEEEproc} to define forward-backward probabilities,
\begin{equation}
\begin{split}
\alpha_{i,t,k,h} 
& = P[(x_{i,1},\ldots,x_{i,t})\trans,s_{i,t,h} = 1\mid z_{i,k} = 1,\boldsymbol{\Theta}^{(j)}], \\
\beta_{i,t,k,h} 
& = P[(x_{i,t+1},\ldots,x_{i,T})\trans,s_{i,t,h} = 1\mid z_{i,k} = 1,\boldsymbol{\Theta}^{(j)}].
\nonumber
\end{split}
\end{equation}
The quantities that need to be computed in the E-step are listed below:
\begin{equation*}
\begin{split}
\tau_{i,k}(\boldsymbol{\Theta}^{(j)}) & = E[I(z_{i,k}=1)\mid\boldsymbol{x}_i,\boldsymbol{\Theta}^{(j)}], \\
\xi_{i,t,k,p,q}(\boldsymbol{\Theta}^{(j)}) & = E[I(s_{i,t-1,p}=1, s_{i,t,q}) = 1\mid z_{i,k} = 1,\boldsymbol{x}_i,\boldsymbol{\Theta}^{(j)}],\\
\gamma_{i,t,k,h}(\boldsymbol{\Theta}^{(j)}) & = E[I(s_{i,t,h}=1)\mid z_{i,k} = 1,\boldsymbol{x}_i,\boldsymbol{\Theta}^{(j)}],\\
\eta_{i,t,h}(\boldsymbol{\Theta}^{(j)}) & = E[I(s_{i,t,h}=1)\mid \boldsymbol{x}_i,\boldsymbol{\Theta}^{(j)}].
\end{split}
\end{equation*}
Consequently, we have that  
\begin{equation}
\begin{split}
Q(\boldsymbol{\Theta}\mid \boldsymbol{\Theta}^{(j)})
= & \sum_{i=1}^N \sum_{k=1}^K \tau_{i,k}(\boldsymbol{\Theta}^{(j)}) \log( \pi_{k,\boldsymbol{s}_{i,1}}) \\
& + \sum_{i=1}^N \sum_{k=1}^K \tau_{i,k}(\boldsymbol{\Theta}^{(j)}) \{\sum_{t=2}^T\sum_{p=1}^M\sum_{q=1}^M \xi_{i,t,k,p,q}(\boldsymbol{\Theta}^{(j)}) \log(\boldsymbol{\Pi}_k)_{p,q}\} \\ 
& + \sum_{i=1}^N \sum_{k=1}^K \tau_{i,k}(\boldsymbol{\Theta}^{(j)}) \{ \sum_{t=1}^T \sum_{p=1}^M \gamma_{i,t,k,p}(\boldsymbol{\Theta}^{(j)}) \log f(x_{i,t} \mid s_{i,t,p} = 1) \}.
\nonumber
\end{split}
\end{equation}

The M-step then maximizes $Q(\boldsymbol{\Theta}\mid \boldsymbol{\Theta}^{(j)})$ to update the parameters, which leads to 

\begin{equation}
\begin{split}
\delta_k^{(j+1)} & = \frac{\sum_{i=1}^N \tau_{i,k}(\boldsymbol{\Theta}^{(j)})}{N},\\
\boldsymbol{\pi}_{k,h}^{(j+1)} & =  \frac{\sum_{i=1}^N \tau_{i,k}(\boldsymbol{\Theta}^{(j)})\gamma_{i,1,k,h}(\boldsymbol{\Theta}^{(j)})}{\sum_{i=1}^N \tau_{i,k}(\boldsymbol{\Theta}^{(j)})},\\
(\boldsymbol{\Pi}_k)_{p,q}^{(j+1)} & = \frac{ \sum_{i=1}^N \tau_{i,k}(\boldsymbol{\Theta}^{(j)}) \{\sum_{t=2}^T \xi_{i,t,k,p,q}(\boldsymbol{\Theta}^{(j)})\} }{\sum_{i=1}^N \tau_{i,k}(\boldsymbol{\Theta}^{(j)}) \{\sum_{t=2}^T \sum_{q'=1}^M \xi_{i,t,k,p,q'}(\boldsymbol{\Theta}^{(j)})\}
},\\
(\epsilon_h^{(r)})^{(j+1)} & = 
\frac{\sum_{i=1}^N \sum_{k=1}^K \tau_{i,k}(\boldsymbol{\Theta}^{(j)}) \sum_{t=1}^T \gamma_{i,t,k,h}(\boldsymbol{\Theta}^{(j)}) I(x_{i,t} = r)}
{\sum_{i=1}^N \sum_{k=1}^K \tau_{i,k}(\boldsymbol{\Theta}^{(j)}) \sum_{t=1}^T \gamma_{i,t,k,h}(\boldsymbol{\Theta}^{(j)})} 
\quad \text{for } r \in \{0, 1\},\\
(a_h^{(j+1)},b_h^{(j+1)}) & = \argmax_{(a_h, b_h)} \sum_{i=1}^N \sum_{k=1}^K \tau_{i,k}(\boldsymbol{\Theta}^{(j)}) \{ \sum_{t=1}^T \gamma_{i,t,k,h}(\boldsymbol{\Theta}^{(j)})\}.
\end{split}
\label{Mstepab}
\end{equation}
The last problem in \eqref{Mstepab} can be solved by a Newton-Raphson algorithm. 

In the above we have mainly focused on the key structures of the algorithm; all the details are provided in Section A 
of the Supplementary Material.

\subsubsection{Initial Values}

Due to the non-convex nature of the problem, the EM algorithm may be sensitive to initial values. We consider two strategies, i.e., the initial values can either be randomly generated or constructed from single-dot analysis. For the latter, specifically, we fit a 0/1 inflated hidden Markov model for each QD, and then apply the K-means algorithm on their estimated transition matrices (with proper alignment) to find the initial clustering pattern of the samples. In practice, we find that combining the multiple random start strategy with the single-dot based initialization leads to the most stable results. In our numerical studies, unless otherwise noted, we fit the model with 10 sets of random initial values as well as the initial values from the single-dot analysis, and the final solution is the one that leads to the highest likelihood value.

\subsubsection{Model Selection \& Inference}

To select the number of clusters and the number of states, several information criteria are available, including AIC \citep{Akaike1974_IEEE}, BIC \citep{Schwarz1978}, and Integrated Completed Likelihood (ICL) \citep{ICL}. Our empirical study shows that these criteria perform well in general, and, as expected, AIC tends to favor a more complex model than BIC and ICL. In practice, we advocate combining the results from several information criteria, using domain knowledge, and examining the clustering patterns of varying numbers of clusters to reach a final conclusion. To obtain the standard errors of the estimated parameters, we utilize the SEM algorithm in \citet{Meng1991}. Details are provided in Section B 
of the Supplementary Material.

\section{Simulation Study}
\label{sec:4}

\subsection{Simulation Setup}

We simulate data from the MHMM-$\beta$ model with $M$ states and $K$ clusters. First, the QDs are randomly allocated to different clusters with mixing probabilities $\delta_k, k \in \{1,\ldots,K\} $, where $\sum_{k=1}^K \delta_k = 1$. For each dot in cluster $k$, we generate its hidden-state sequence as a Markov chain, with the transition matrix $\boldsymbol{\Pi}_k$, $k \in \{1,\ldots,K\}$. We then simulate the intensities given the hidden state sequence by sampling from the 0/1 inflated-Beta distribution with parameters $(a_h,b_h,\epsilon_h^{(0)},\epsilon_h^{(1)})$, $h \in \{1,\ldots, M\}$.

Here, we mainly focus on simulation settings that mimic the QD application. The total number of dots is set to $N = 100$, the number of states is set to $M=3$, and the number of clusters is set to $K =3$. We consider various lengths of the individual QD series, i.e., $T\in \{250,500, 1000, 2000\}$. The mixing probabilities have two scenarios: a relatively balanced partition with $(\delta_1,\delta_2,\delta_3)= (0.3, 0.3, 0.4)$ and an unbalanced partition with $(\delta_1,\delta_2,\delta_3) = (0.7, 0.2, 0.1)$. 

We consider several scenarios for transition matrices and state distributions.

\noindent \textbf{Scenario 1}. In the first scenario, we closely mimic the estimated MHMM-$\beta$ model from the real data application. The true transition matrices are 
\begin{equation}
\boldsymbol{\Pi}_1=
\begin{bmatrix} 
0.848 & 0.127 & 0.025 \\ 
0.060 & 0.794 & 0.146 \\ 
0.017 & 0.186 & 0.796 
\end{bmatrix},
\boldsymbol{\Pi}_2=
\begin{bmatrix} 
0.795 & 0.205 & 0.000 \\
0.000 & 0.994 & 0.006 \\ 
0.000 & 0.001 & 0.999 
\end{bmatrix},
\boldsymbol{\Pi}_3=
\begin{bmatrix}
0.938 & 0.057 & 0.005 \\ 
0.013 & 0.924 & 0.063 \\ 
0.001 & 0.067 & 0.932 
\end{bmatrix},
\nonumber
\end{equation}
and the parameters for the three states are set as 
\begin{align*}
(a_1,b_1,\epsilon_1^{(0)},\epsilon_1^{(1)}) & = (2.195, 5.183, 0.025, 0.000),\\
(a_2,b_2,\epsilon_2^{(0)},\epsilon_2^{(1)}) & = (10.077, 6.805, 0.000, 0.001),\\
(a_3,b_3,\epsilon_3^{(0)},\epsilon_3^{(1)}) & = (11.658, 3.227, 0.000, 0.017).
\end{align*}

\noindent \textbf{Scenario 2}. In the second scenario, we modify the above settings to make the three clusters more distinct and the proportions of 0/1 inflation larger. The transition matrices are given by
\begin{equation}
\boldsymbol{\Pi}_1=
\begin{bmatrix} 
0.500 & 0.250 & 0.250 \\ 
0.250 & 0.500 & 0.250 \\ 
0.250 & 0.250 & 0.500 
\end{bmatrix}
,
\boldsymbol{\Pi}_2=
\begin{bmatrix} 
0.940 & 0.050 & 0.010 \\
0.010 & 0.920 & 0.070 \\ 
0.010 & 0.050 & 0.940 
\end{bmatrix}
,
\boldsymbol{\Pi}_3=
\begin{bmatrix}
0.840 & 0.120 & 0.004 \\ 
0.060 & 0.730 & 0.210 \\ 
0.020 & 0.170 & 0.810 
\end{bmatrix},
\nonumber
\end{equation}
and the parameters for the three states are set as 
\begin{align*}
(a_1,b_1,\epsilon_1^{(0)},\epsilon_1^{(1)}) &= (2.000, 4.000, 0.100, 0.010),\\ 
(a_2,b_2,\epsilon_2^{(0)},\epsilon_2^{(1)}) &= (8.000, 4.000, 0.050, 0.050),\\ 
(a_3,b_3,\epsilon_3^{(0)},\epsilon_3^{(1)}) &= (10.000, 2.000, 0.010, 0.100).
\end{align*}

The two scenarios differ primarily in how distinct the clusters are. In Scenario 1, the long-run probabilities (\(\boldsymbol{\pi}_1 = (0.213, 0.443, 0.344)\), \(\boldsymbol{\pi}_2 = (0.000, 0.143, 0.857)\), \(\boldsymbol{\pi}_3 = (0.104, 0.461, 0.435)\)) indicate that Clusters 1 and 3 transit among the three states with similar probabilities, making them more challenging to separate. In contrast, in Scenario 2, the long-run probabilities for all three clusters (\(\boldsymbol{\pi}_1 = (0.333, 0.333, 0.333)\), \(\boldsymbol{\pi}_2 = (0.143, 0.385, 0.473)\), \(\boldsymbol{\pi}_3 = (0.192, 0.365, 0.443)\)) suggest comparatively sharper cluster distinctions. 

\subsection{Competing Methods}
We compare our proposed MHMM-$\beta$ methods with several competing methods. Its closest competitor is the Gaussian mixture HMM model, denoted as MHMM-G, which jointly analyzes all the QDs but uses Gaussian state distributions. 

Another set of competitors is based on single-dot analysis, including the Gaussian HMM model (HMM-G), and the 0/1 inflated-Beta HMM model (HMM-$\beta$). In order to perform clustering via single-dot analysis, the K-means algorithm is applied to the estimated individual transition matrices. The transition matrices within each cluster are estimated by exponentiating the average of the log-transformed individual transition matrices \citep{McKinney2006_Biophy}. 

Besides the above methods, we also include an oracle procedure, denoted as MHMM-$\beta^*$, as a benchmark, where the clustering information is assumed to be known and the true parameter values are used as initialization.

\subsection{Evaluation Metrics}

To compare model estimation, we report the error in estimating the means of the state distributions, i.e., $\mbox{Er}(\widehat{\mu}) = \sqrt{\sum_{h=1}^{M}(\widehat{\mu}_h - \mu_h)^2}$, the error in estimating the variances of the state distributions, i.e., $\mbox{Er}(\widehat{\sigma}^2) = \sqrt{\sum_{h=1}^{M} (\widehat{\sigma}_h^2 - \sigma_h^2)^2}$, the error in estimating the mixing probabilities, i.e., $\mbox{Er}(\widehat{\delta}) = \sqrt{\sum_{k=1}^K (\widehat{\delta}_k- \delta_k)^2}$, and the error in estimating the transition matrices, i.e., $\mbox{Er}(\widehat{\boldsymbol{\Pi}}) = \sum_{k=1}^K \Vert \boldsymbol{\widehat{\Pi}}_k - \boldsymbol{\Pi}_k \Vert_F$. Here $\widehat{\mu}_h$, $\widehat{\sigma}_h^2$, $\widehat{\delta}_k$,  and $\boldsymbol{\widehat{\Pi}}_k$ are the estimates of their true counterparts, $\mu_h$, $\sigma_h^2$, $\delta_k$, and $\boldsymbol{\Pi}_k$, respectively. For MHMM-$\beta$ and MHMM-$\beta^*$, we also report the estimation error of the inflated-Beta distribution parameters, i.e., $\mbox{Er}(\widehat{\boldsymbol{\theta}}) = \sqrt{\sum_{h=1}^M \Vert \boldsymbol{\widehat{\theta}}_h-\boldsymbol{\theta}_h \Vert_2^2}$, with 
$\widehat{\boldsymbol{\theta}}_h = (\widehat{a}_h, \widehat{b}_h,\widehat{\epsilon}_h^{(0)},\widehat{\epsilon}_h^{(1)})\trans$ and $\boldsymbol{\theta}_h = (a_h, b_h,\epsilon_h^{(0)},\epsilon_h^{(1)})\trans$ representing the estimated and true parameters, respectively.

To measure the accuracy of retrieving the true cluster pattern of the dots, we report the percentage of correct clustering: $\mbox{CC} = \sum_{i=1}^N I(\widehat{\boldsymbol{z}}_i = \boldsymbol{z}_i)/N$. 

The simulation under each setting is replicated 100 times and the results are averaged. The potential label switching problem is solved by the Hungarian Method in linear sum assignment problem \citep{Hungarian}.

\subsection{Simulation Results}

Tables \ref{table:scenario1_length_250_500}–\ref{table:scenario2_length_1000_2000} present simulation results under two different scenarios, each tested with balanced and unbalanced cluster partitions. The oracle procedure, MHMM-$\beta^*$, provides a benchmark for parameter estimation by assuming that the cluster memberships are known and the true estimates are provided as initials. As expected, it achieves the lowest estimation errors, offering insight into how well the proposed method and its competitors perform.

Across all simulation settings, the proposed MHMM-$\beta$ method performs well in parameter estimation and clustering accuracy. In many cases, the estimation errors for MHMM-$\beta$ come close to those attained by the oracle method. In both balanced and unbalanced scenarios, MHMM-$\beta$ outperforms the single-dot methods (HMM-G and HMM-$\beta$), which do not pool information across QDs. These single-dot approaches tend to have higher errors in transition matrix estimation and yield lower accuracy in retrieving the true cluster structure. The MHMM-G method partially captures some shared dynamics among QDs but is also consistently outperformed by MHMM-$\beta$. This difference emphasizes the value of an inflated-Beta emission distribution for modeling the standardized intensities and handling zero/one inflation features that cannot be adequately addressed by Gaussian assumptions. 

When comparing Scenario 1 vs. 2 and balanced vs. unbalanced partitions, the settings of more overlapping clusters and unbalanced partitions pose more challenges in identifying and estimating smaller clusters, leading to slightly higher errors and lower clustering accuracy; nonetheless, MHMM-$\beta$ remains robust and still outperforms its competitors under these conditions. Across all methods, longer time series contribute to more accurate parameter estimates and higher clustering accuracy.

Overall, the simulation studies confirm that MHMM-$\beta$ successfully captures both the hidden-state dynamics and the QD clustering patterns. By aligning the state/emission distributions with the physical and experimental characteristics of the QDs, the proposed method offers considerable advantages in parameter estimation and clustering. 

Additional simulation results are reported in Section C of the Supplemental Materials. In particular, we have conducted simulation studies to examine the performance of several information criteria, including the Akaike Information Criterion (AIC) \citep{Akaike1974_IEEE}, the Bayesian Information Criterion (BIC) \citep{Schwarz1978}, and the Integrated Completed Likelihood (ICL) criterion \citep{ICL}, on selecting the number of clusters. Our results illustrate the well-known differences in how each criterion penalizes model complexity and cluster separation; while they often converge on the truth, discrepancies sometimes arise in cases of overlapping or unevenly sized clusters. Therefore, in practice, we advocate the examination of the cluster patterns of different models and the use of both information criteria and domain knowledge for model selection.

\begin{table}[H]
  \centering
  \scriptsize
\caption{Simulation: Results for Scenario 1 with balanced partition. For better presentation, values in $\mbox{Er}(\widehat{\boldsymbol{\delta}})$ and $\mbox{Er}(\widehat{\boldsymbol{\theta}})$ are multiplied by $10$, values in $\mbox{Er}(\widehat{\mu})$ are multiplied by $10^2$, and values in $\mbox{Er}(\widehat{\sigma}^2)$ are multiplied by $10^3$.}
\label{table:scenario1_length_250_500}
\makebox[\textwidth][c]{
\begin{tabular}{ l c | c c c | c c c}
  Method & Length ($T$) &  $\mbox{Er}(\widehat{{\mu}})$ & $\mbox{Er}(\widehat{\sigma}^2)$ & $\mbox{Er}(\widehat{\boldsymbol{\delta}})$ & $\mbox{Er}(\widehat{\boldsymbol{\theta}})$ & $\mbox{Er}(\widehat{\boldsymbol{\Pi}})$ & \mbox{CC} \\
  \hline
  \multicolumn{8}{c}{Balanced, Sample size ($N$) = 100}\\
  \hline
MHMM-$\beta^*$ &  \multirow{5}{*}{250} & 0.36 (0.02) & 0.77 (0.05) & 0.00 (0.00) & 3.17 (0.13) & 0.13 (0.01) &  \\ 
  HMM-G &   & 14.37 (0.11) & 9.95 (0.06) & 3.21 (0.09) &  & 2.07 (0.02) & 0.53 (0.00) \\ 
  HMM-$\beta$ &   & 20.61 (0.15) & 9.73 (0.08) & 3.78 (0.07) &  & 2.15 (0.03) & 0.53 (0.00) \\ 
  MHMM-G &   & 0.74 (0.03) & 2.19 (0.08) & 1.13 (0.13) &  & 0.62 (0.06) & 0.89 (0.01) \\ 
  MHMM-$\beta$ &   & 0.38 (0.02) & 0.79 (0.05) & 0.60 (0.11) & 3.26 (0.13) & 0.36 (0.05) & 0.94 (0.01) \\ 
  \hline
MHMM-$\beta^*$ & \multirow{5}{*}{500}  & 0.27 (0.01) & 0.53 (0.03) & 0.00 (0.00) & 2.08 (0.09) & 0.11 (0.01) &  \\ 
  HMM-G &   & 13.68 (0.09) & 8.35 (0.04) & 3.72 (0.05) &  & 2.08 (0.02) & 0.52 (0.00) \\ 
  HMM-$\beta$ &   & 19.88 (0.13) & 6.73 (0.06) & 3.74 (0.05) &  & 1.99 (0.02) & 0.53 (0.00) \\ 
  MHMM-G &   & 0.78 (0.03) & 2.25 (0.06) & 1.70 (0.16) &  & 0.83 (0.07) & 0.86 (0.01) \\ 
  MHMM-$\beta$ &   & 0.28 (0.01) & 0.55 (0.03) & 0.80 (0.15) & 2.36 (0.10) & 0.42 (0.06) & 0.93 (0.01) \\ 
  \hline  
MHMM-$\beta^*$ &  \multirow{5}{*}{1000} & 0.19 (0.01) & 0.40 (0.02) & 0.00 (0.00) & 1.58 (0.06) & 0.11 (0.01) &  \\ 
  HMM-G &   & 12.79 (0.08) & 7.51 (0.04) & 3.60 (0.03) &  & 2.03 (0.01) & 0.57 (0.00) \\ 
  HMM-$\beta$ &   & 18.69 (0.11) & 5.11 (0.05) & 3.79 (0.02) &  & 1.95 (0.01) & 0.58 (0.00) \\ 
  MHMM-G &   & 0.65 (0.02) & 2.16 (0.04) & 1.56 (0.17) &  & 0.70 (0.06) & 0.87 (0.01) \\ 
  MHMM-$\beta$ &   & 0.22 (0.01) & 0.43 (0.02) & 0.58 (0.14) & 1.84 (0.08) & 0.34 (0.05) & 0.96 (0.01) \\ 
  \hline
MHMM-$\beta^*$ & \multirow{5}{*}{2000} & 0.14 (0.01) & 0.26 (0.01) & 0.00 (0.00) & 1.12 (0.04) & 0.09 (0.01) &  \\ 
  HMM-G &   & 11.81 (0.04) & 7.18 (0.02) & 3.71 (0.04) &  & 2.01 (0.02) & 0.61 (0.01) \\ 
  HMM-$\beta$ &   & 17.73 (0.07) & 4.25 (0.03) & 3.91 (0.01) &  & 1.95 (0.01) & 0.61 (0.00) \\ 
  MHMM-G &   & 0.67 (0.02) & 2.18 (0.03) & 1.25 (0.16) &  & 0.61 (0.06) & 0.91 (0.01) \\ 
  MHMM-$\beta$ &   & 0.16 (0.01) & 0.35 (0.02) & 0.69 (0.15) & 1.51 (0.07) & 0.43 (0.06) & 0.95 (0.01) \\
  \hline
  \end{tabular} 
}
\end{table}

\begin{table}[H]
  \centering
  \scriptsize
\caption{Simulation: Results for Scenario 1 with unbalanced partition. For better presentation, values in $\mbox{Er}(\widehat{\boldsymbol{\delta}})$ and $\mbox{Er}(\widehat{\boldsymbol{\theta}})$ are multiplied by $10$, values in $\mbox{Er}(\widehat{\mu})$ are multiplied by $10^2$, and values in $\mbox{Er}(\widehat{\sigma}^2)$ are multiplied by $10^3$.}
\label{table:scenario1_length_1000_2000}
\makebox[\textwidth][c]{
\begin{tabular}{ l c | c c c | c c c}
  Method & Length ($T$) &  $\mbox{Er}(\widehat{{\mu}})$ & $\mbox{Er}(\widehat{\sigma}^2)$ & $\mbox{Er}(\widehat{\boldsymbol{\delta}})$ & $\mbox{Er}(\widehat{\boldsymbol{\theta}})$ & $\mbox{Er}(\widehat{\boldsymbol{\Pi}})$ & \mbox{CC} \\
  \hline
  \multicolumn{8}{c}{Unbalanced, Sample size ($N$) = 100}\\
  \hline
MHMM-$\beta^*$ & \multirow{5}{*}{250} & 0.37 (0.02) & 0.66 (0.04) & 0.00 (0.00) & 3.35 (0.15) & 0.13 (0.01) &  \\ 
  HMM-G &   & 8.33 (0.09) & 8.26 (0.06) & 1.34 (0.08) &  & 2.08 (0.02) & 0.76 (0.01) \\ 
  HMM-$\beta$ &   & 14.62 (0.14) & 7.12 (0.08) & 1.25 (0.04) &  & 2.20 (0.02) & 0.80 (0.00) \\ 
  MHMM-G &   & 1.25 (0.03) & 3.46 (0.06) & 1.74 (0.13) &  & 1.10 (0.07) & 0.84 (0.01) \\ 
  MHMM-$\beta$ &   & 0.38 (0.02) & 0.69 (0.04) & 1.92 (0.16) & 3.74 (0.15) & 0.49 (0.05) & 0.85 (0.01) \\ 
  \hline
MHMM-$\beta^*$ & \multirow{5}{*}{500} & 0.26 (0.01) & 0.48 (0.03) & 0.00 (0.00) & 2.24 (0.10) & 0.12 (0.01) &  \\ 
  HMM-G &   & 7.97 (0.05) & 7.42 (0.04) & 1.31 (0.02) &  & 2.15 (0.02) & 0.82 (0.00) \\ 
  HMM-$\beta$ &   & 13.98 (0.12) & 4.71 (0.06) & 1.45 (0.03) &  & 2.17 (0.02) & 0.81 (0.00) \\ 
  MHMM-G &   & 1.14 (0.03) & 3.39 (0.06) & 1.15 (0.11) &  & 0.93 (0.07) & 0.89 (0.01) \\ 
  MHMM-$\beta$ &   & 0.27 (0.01) & 0.53 (0.03) & 0.98 (0.14) & 2.63 (0.13) & 0.38 (0.05) & 0.93 (0.01) \\ 
  \hline   
MHMM-$\beta^*$ & \multirow{5}{*}{1000} & 0.21 (0.01) & 0.39 (0.02) & 0.00 (0.00) & 1.69 (0.06) & 0.14 (0.02) &  \\ 
  HMM-G &   & 7.76 (0.04) & 7.01 (0.03) & 1.30 (0.02) &  & 2.19 (0.02) & 0.83 (0.00) \\ 
  HMM-$\beta$ &   & 13.43 (0.09) & 3.57 (0.04) & 1.39 (0.02) &  & 2.17 (0.02) & 0.83 (0.00) \\ 
 MHMM-G &   & 1.11 (0.02) & 3.38 (0.05) & 1.04 (0.10) &  & 0.77 (0.06) & 0.91 (0.01) \\ 
  MHMM-$\beta$ &   & 0.22 (0.01) & 0.43 (0.02) & 0.47 (0.10) & 2.15 (0.11) & 0.27 (0.03) & 0.97 (0.01) \\ 
  \hline 
MHMM-$\beta^*$ & \multirow{5}{*}{2000} & 0.13 (0.01) & 0.26 (0.01) & 0.00 (0.00) & 1.14 (0.05) & 0.12 (0.01) &  \\ 
  HMM-G &   & 7.63 (0.03) & 6.79 (0.02) & 1.22 (0.02) &  & 2.17 (0.02) & 0.84 (0.00) \\ 
  HMM-$\beta$ &   & 12.97 (0.07) & 3.08 (0.03) & 1.33 (0.01) &  & 2.17 (0.01) & 0.84 (0.00) \\ 
  MHMM-G &   & 1.07 (0.02) & 3.37 (0.03) & 0.64 (0.11) &  & 0.45 (0.04) & 0.95 (0.01) \\ 
  MHMM-$\beta$ &   & 0.15 (0.01) & 0.39 (0.02) & 0.24 (0.08) & 2.21 (0.10) & 0.20 (0.02) & 0.98 (0.01) \\ 
  \hline
  \end{tabular}
}  
\end{table}

\begin{table}[H]
  \centering
  \scriptsize
\caption{Simulation: Results for Scenario 2 with balanced partition. For better presentation, values in $\mbox{Er}(\widehat{\boldsymbol{\delta}})$ and $\mbox{Er}(\widehat{\boldsymbol{\theta}})$ are multiplied by $10$, and values in $\mbox{Er}(\widehat{\mu})$ and $\mbox{Er}(\widehat{\sigma}^2)$ are multiplied by $10^2$.}
\label{table:scenario2_length_250_500}
\makebox[\textwidth][c]{
\begin{tabular}{ l c | c c c | c c c}
  Method & Length ($T$) &  $\mbox{Er}(\widehat{{\mu}})$ & $\mbox{Er}(\widehat{\sigma}^2)$ & $\mbox{Er}(\widehat{\boldsymbol{\delta}})$ & $\mbox{Er}(\widehat{\boldsymbol{\theta}})$ & $\mbox{Er}(\widehat{\boldsymbol{\Pi}})$ & \mbox{CC} \\
  \hline
  \multicolumn{8}{c}{Balanced, Sample size ($N$) = 100}\\
  \hline
MHMM-$\beta^*$ & \multirow{5}{*}{250} & 0.58 (0.03) & 2.25 (0.11) & 0.00 (0.00) & 3.31 (0.13) & 0.10 (0.00) &  \\ 
  HMM-G &   & 6.52 (0.04) & 32.94 (0.05) & 1.05 (0.06) &  & 0.98 (0.01) & 0.75 (0.01) \\ 
  HMM-$\beta$ &   & 5.81 (0.26) & 9.01 (0.18) & 1.43 (0.09) &  & 0.78 (0.04) & 0.73 (0.01) \\ 
  MHMM-G &   & 6.79 (0.08) & 30.73 (0.09) & 0.79 (0.07) &  & 1.00 (0.02) & 0.79 (0.01) \\ 
  MHMM-$\beta$ &   & 0.92 (0.05) & 2.92 (0.15) & 0.38 (0.04) & 7.95 (0.51) & 0.22 (0.01) & 0.97 (0.00) \\ 
  \hline
MHMM-$\beta^*$ & \multirow{5}{*}{500} & 0.40 (0.02) & 1.54 (0.06) & 0.00 (0.00) & 2.27 (0.09) & 0.07 (0.00) &  \\ 
  HMM-G &   & 6.65 (0.03) & 31.94 (0.03) & 0.86 (0.06) &  & 0.89 (0.01) & 0.83 (0.00) \\ 
  HMM-$\beta$ &   & 2.86 (0.09) & 6.84 (0.16) & 0.50 (0.03) &  & 0.32 (0.01) & 0.90 (0.00) \\ 
  MHMM-G &   & 6.71 (0.07) & 30.75 (0.07) & 0.57 (0.07) &  & 0.94 (0.01) & 0.88 (0.01) \\ 
  MHMM-$\beta$ &   & 0.70 (0.04) & 2.21 (0.12) & 0.21 (0.08) & 6.28 (0.39) & 0.19 (0.02) & 0.98 (0.01) \\  
  \hline
MHMM-$\beta^*$ & \multirow{5}{*}{1000} & 0.28 (0.01) & 1.03 (0.05) & 0.00 (0.00) & 1.73 (0.08) & 0.05 (0.00) &  \\ 
  HMM-G &   & 6.72 (0.02) & 31.35 (0.03) & 0.83 (0.06) &  & 0.88 (0.01) & 0.90 (0.00) \\ 
  HMM-$\beta$ &   & 1.60 (0.05) & 4.95 (0.11) & 0.35 (0.02) &  & 0.23 (0.00) & 0.96 (0.00) \\ 
  MHMM-G &   & 6.54 (0.06) & 30.79 (0.07) & 0.25 (0.05) &  & 0.91 (0.01) & 0.97 (0.01) \\ 
  MHMM-$\beta$ &   & 0.84 (0.07) & 2.31 (0.17) & 0.01 (0.01) & 7.70 (0.64) & 0.16 (0.01) & 1.00 (0.00) \\ 
  \hline
MHMM-$\beta^*$ & \multirow{5}{*}{2000} & 0.19 (0.01) & 0.75 (0.03) & 0.00 (0.00) & 1.21 (0.05) & 0.04 (0.00) &  \\ 
  HMM-G &   & 6.79 (0.01) & 30.97 (0.02)  & 0.54 (0.03) &  & 0.89 (0.00) & 0.96 (0.00) \\ 
  HMM-$\beta$ &   & 1.00 (0.02) & 3.96 (0.09) & 0.07 (0.01) &  & 0.21 (0.00) & 0.99 (0.00) \\ 
  MHMM-G &   & 6.50 (0.06) & 30.68 (0.07) & 0.02 (0.01) &  & 0.88 (0.01) & 1.00 (0.00) \\ 
  MHMM-$\beta$ &   & 0.59 (0.04) & 1.78 (0.13) & 0.08 (0.06) & 5.70 (0.48) & 0.15 (0.02) & 0.99 (0.01) \\ 
  \hline   
  \end{tabular} 
}
\end{table}

\begin{table}[H]
  \centering
  \scriptsize
\caption{Simulation: Results for Scenario 2 with unbalanced partition. For better presentation, values in $\mbox{Er}(\widehat{\boldsymbol{\delta}})$ and $\mbox{Er}(\widehat{\boldsymbol{\theta}})$ are multiplied by $10$, and values in $\mbox{Er}(\widehat{\mu})$ and $\mbox{Er}(\widehat{\sigma}^2)$ are multiplied by $10^2$.}
\label{table:scenario2_length_1000_2000}
\makebox[\textwidth][c]{
\begin{tabular}{ l c | c c c | c c c}
  Method & Length ($T$) &  $\mbox{Er}(\widehat{{\mu}})$ & $\mbox{Er}(\widehat{\sigma}^2)$ & $\mbox{Er}(\widehat{\boldsymbol{\delta}})$ & $\mbox{Er}(\widehat{\boldsymbol{\theta}})$ & $\mbox{Er}(\widehat{\boldsymbol{\Pi}})$ & \mbox{CC} \\
  \hline 
  \multicolumn{8}{c}{Unbalanced, Sample size ($N$) = 100}\\
  \hline
MHMM-$\beta^*$ & \multirow{5}{*}{250} & 0.69 (0.03) & 2.64 (0.11) & 0.00 (0.00) & 3.51 (0.17) & 0.12 (0.00) &  \\ 
  HMM-G &   & 6.07 (0.03) & 33.44 (0.04) & 2.05 (0.13) &  & 1.24 (0.03) & 0.74 (0.01) \\ 
  HMM-$\beta$ &   & 4.88 (0.24) & 11.06 (0.25) & 3.21 (0.07) &  & 1.18 (0.01) & 0.61 (0.01) \\ 
  MHMM-G &   & 7.08 (0.07) & 32.23 (0.06) & 2.57 (0.17) &  & 1.27 (0.03) & 0.74 (0.02) \\ 
  MHMM-$\beta$ &   & 1.28 (0.08) & 4.37 (0.22) & 2.39 (0.19) & 9.09 (0.48) & 0.67 (0.03) & 0.76 (0.02) \\
  \hline  
  MHMM-$\beta^*$ & \multirow{5}{*}{500} & 0.51 (0.03) & 1.88 (0.08) & 0.00 (0.00) & 2.87 (0.13) & 0.08 (0.00) &  \\ 
  HMM-G &   & 6.55 (0.02) & 32.47 (0.03) & 1.14 (0.08) &  & 0.89 (0.02) & 0.89 (0.01) \\ 
  HMM-$\beta$ &   & 2.81 (0.07) & 9.91 (0.17) & 2.35 (0.12) &  & 0.92 (0.03) & 0.72 (0.01) \\ 
  MHMM-G &   & 7.09 (0.05) & 32.03 (0.05) & 1.71 (0.17) &  & 1.10 (0.03) & 0.82 (0.02) \\ 
  MHMM-$\beta$ &   & 1.04 (0.07) & 3.33 (0.19) & 2.16 (0.19) & 7.60 (0.41) & 0.61 (0.04) & 0.78 (0.02) \\ 
  \hline 
MHMM-$\beta^*$ & \multirow{5}{*}{1000} & 0.35 (0.01) & 1.41 (0.05) & 0.00 (0.00) & 1.98 (0.08) & 0.06 (0.00) &  \\ 
  HMM-G &   & 6.83 (0.02) & 31.93 (0.02) & 0.82 (0.04) &  & 0.85 (0.01) & 0.93 (0.00) \\ 
  HMM-$\beta$ &   & 2.03 (0.04) & 9.08 (0.13) & 0.37 (0.06) &  & 0.26 (0.01) & 0.96 (0.01) \\ 
  MHMM-G &   & 6.98 (0.05) & 32.04 (0.04) & 1.57 (0.18) &  & 1.09 (0.03) & 0.85 (0.02) \\ 
  MHMM-$\beta$ &   & 0.77 (0.05) & 2.56 (0.15) & 0.80 (0.15) & 7.19 (0.34) & 0.31 (0.03) & 0.92 (0.02) \\ 
  \hline 
MHMM-$\beta^*$ & \multirow{5}{*}{2000} & 0.24 (0.01) & 0.93 (0.04) & 0.00 (0.00) & 1.40 (0.06) & 0.04 (0.00) &  \\ 
  HMM-G &   & 6.97 (0.01) & 31.67 (0.01) & 0.48 (0.03) &  & 0.87 (0.01) & 0.97 (0.00) \\ 
  HMM-$\beta$ &   & 1.65 (0.02) & 8.04 (0.10) & 0.06 (0.01) &  & 0.20 (0.00) & 1.00 (0.00) \\ 
  MHMM-G &   & 6.87 (0.05) & 32.04 (0.04) & 0.79 (0.16) &  & 1.02 (0.02) & 0.92 (0.02) \\ 
  MHMM-$\beta$ &   & 0.77 (0.07) & 2.50 (0.21) & 0.30 (0.09) & 6.83 (0.38) & 0.23 (0.03) & 0.97 (0.01) \\ 
  \hline
  \end{tabular}
}  
\end{table}

\section{Case Study: Integrative Analysis of Quantum Dots}
\label{sec:5}
We applied the proposed method to analyze the quantum dot data described in Section \ref{sec:2}. Chemists believe that these quantum dots may transit between up to 3 states, corresponding to relatively low, median, and high levels of intensity of emitting  photons under continuous excitation. Owing to the denosing and standardization procedure, all the dots became comparable. Thus, our main interests were to identify the three intensity states across all quantum dots, and examine cluster patterns of quantum dots with unique transition behaviors across intensity states; in particular, we hope to verify that the dots could exhibit either ``Blinking'' or ``Flickering'' styles of intensity fluctuations.

\subsection{Cluster Patterns} 

\begin{figure}[htp]
  \centering
  \includegraphics[width=0.98\linewidth]{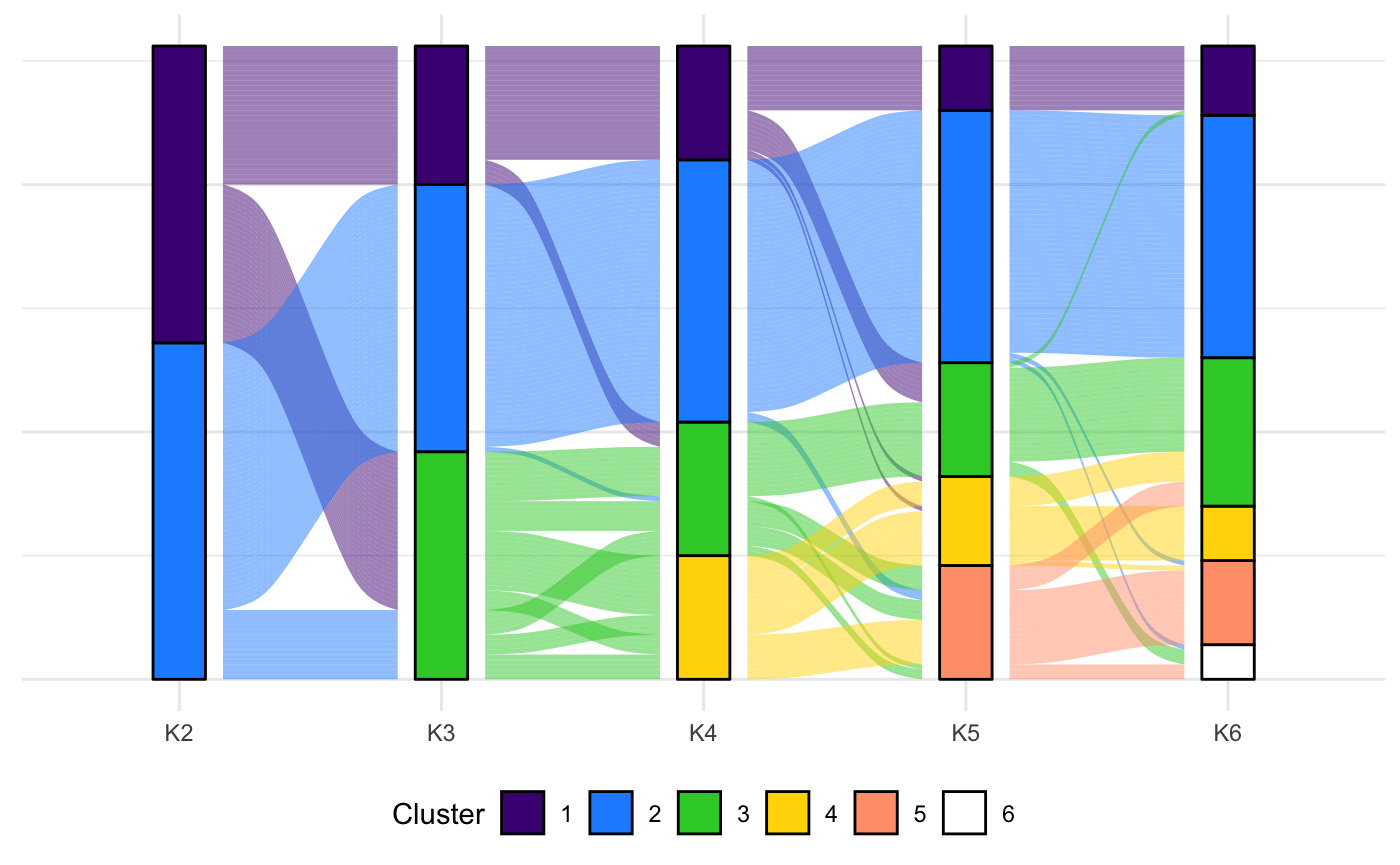}
  \caption{Quantum dot analysis: Cluster patterns.}
  \label{fig:ClusterAssignChange_new}
\end{figure}

We applied the proposed MHMM-$\beta$ approach with varying number of clusters. The cluster patterns are displayed in Figure \ref{fig:ClusterAssignChange_new}, in which the colors or the gray levels represent different clusters and the stripes show how the dots are put into more and more clusters from the left to the right.

Starting from the 2-cluster model, it can be seen that in the 3-cluster model, the third cluster is formed by QDs from both clusters. From there, as the number of clusters increases, the new clusters are mainly formed from further splitting the third cluster, and the first two clusters remain relatively stable throughout. This clear pattern suggests that the 3-cluster model is the most stable and informative. We have also computed AIC, BIC and ICL values for these models and unfortunately they have large discrepancy: AIC and BIC would select a model with more than 3 clusters while ICL suggests 2 clusters. This in fact is consistent with our simulation studies and comparative studies in the literature, which have shown that when the clusters are too overlapped, ICL tends to favor less number of clusters than BIC. From these results and after consulting with our chemistry collaborators, we decided to mainly focus on the results from the three-cluster model; as to be shown below, the results indeed are highly interpretable based on principles of chemistry and physics. We also provide the results from other models in Section D of the Supplementary Material. 

\subsection{Results from 3-Cluster MHMM-$\beta$ Model}

\begin{figure}[H]
  \centering
  \includegraphics[width=0.8\linewidth]{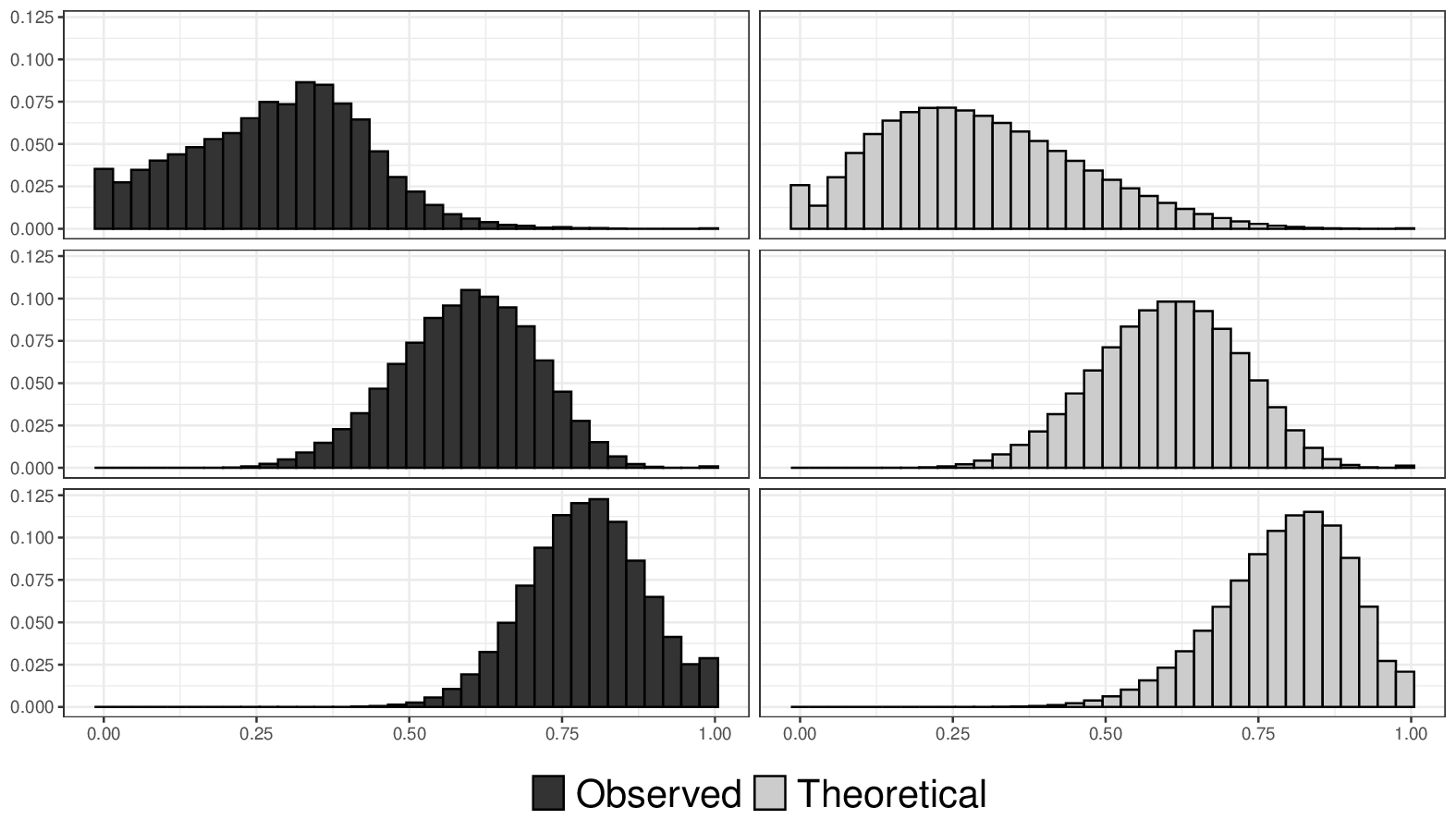}
  \caption{Quantum dot analysis: Comparison of the three intensity states.}
  \label{fig:CompareState}
\end{figure}

In Figure \ref{fig:CompareState}, the right panel shows the estimated 0/1 inflated Beta distributions of the three intensity states from the MHMM-$\beta$ model. With the estimated state series, we obtain the empirical distributions of the three states directly from the standardized intensity data, which are shown in the left panel. As expected, the estimated and the empirical distributions are quite similar to each other, suggesting that here are no apparent outliers, and the MHMM-$\beta$ model fits the data well and successfully captures the collective behaviors of intensity fluctuations of all QDs.

Table \ref{table:resultpbaram} reports the estimation results. Besides the parameter estimates from the inflated Beta distributions, we also report the estimated mean and variance of each distribution, where  
\begin{align*}
\widehat{\mu}_h & =  \widehat{a}_h/(\widehat{a}_h+\widehat{b}_h) + \widehat{\epsilon}_h^{(1)},\\
\widehat{\sigma}_h^2 & = (1 - \widehat{\epsilon}_h^{(0)} - \widehat{\epsilon}_h^{(1)}) \frac{(\widehat{a}_h^3 + \widehat{a}_h^2\widehat{b}_h+\widehat{a}_h^2+\widehat{a}_h\widehat{b}_h)}{(\widehat{a}_h + \widehat{b}_h)^2(\widehat{a}_h + \widehat{b}_h +1)} + \widehat{\epsilon}_h^{(1)},
\end{align*}
for $h=1,\ldots,3$. As seen from both Figure \ref{fig:CompareState} and the estimated parameters in Table \ref{table:resultpbaram}, the three states clearly correspond to relatively low, medium, and high intensity levels. The low-intensity state (State 1) accommodates intensities in the range of about 0.00 to 0.50, with mean 0.290 and relatively the highest variance. The medium-intensity state (State 2) mostly concentrates within the 0.50 to 0.80 spectrum, with mean 0.597 and a relatively small variance. The high-intensity state (State 3) covers the range of about 0.75 to 1.00, with mean 0.787 and relatively the smallest variance. The zero and one inflation rates are also reflected in Figure \ref{fig:CompareState}. The low-intensity state accommodates the zero intensities, while the high-intensity state encompasses all the intensities elevated to one. We remark that because the standardized data contain virtually no ``1'' observations in State 1 and no ``0'' observations in States 2 and 3, the corresponding MLEs for $\epsilon_h^{(1)}$ (State 1) and $\epsilon_h^{(0)}$ (States 2-3) lie on the boundary, and the resulting asymptotic standard errors therefore appear extremely small.

\begin{table}[H]
\centering
\caption{Quantum dot analysis: Estimated parameters of the fitted 3-cluster MHMM-$\beta$ model.}\label{table:resultpbaram}
\begin{tabular}{c c c c}
\hline
& State 1 ($h=1$) & State 2 ($h=2$)& State 3 ($h=3$)\\
$\widehat{\epsilon}_h^{(0)}$ & 0.025 $(4.917\times 10^{-3})$ & 0.000 $(1.276\times 10^{-10})$ & 0.000 $(6.301\times 10^{-15})$ \\
$\widehat{\epsilon}_h^{(1)}$ & 0.000 $(4.439\times 10^{-9})$ & 0.001 $(2.714\times 10^{-3})$ & 0.017 $(9.022\times 10^{-4})$ \\
$\widehat{a}_h$ & 2.195 $(8.580\times 10^{-3})$ & 10.077 $(3.336\times 10^{-2})$ & 11.658 $(1.803\times 10^{-3})$ \\
$\widehat{b}_h$ & 5.183 $(4.178\times 10^{-3})$ & 6.805 $(9.739\times 10^{-4})$ & 3.227 $(1.015 \times 10^{-2})$ \\
$\widehat{\mu}_h$ & 0.290 & 0.597 & 0.787\\
$\widehat{\sigma}_h^2$ & 0.0266 & 0.0137 & 0.0113\\
\hline
\end{tabular}
\end{table}

\subsection{State Transition Patterns and Clusters of Quantum Dots}

\begin{figure}[H]
  \centering
  \includegraphics[width = 0.9\linewidth]{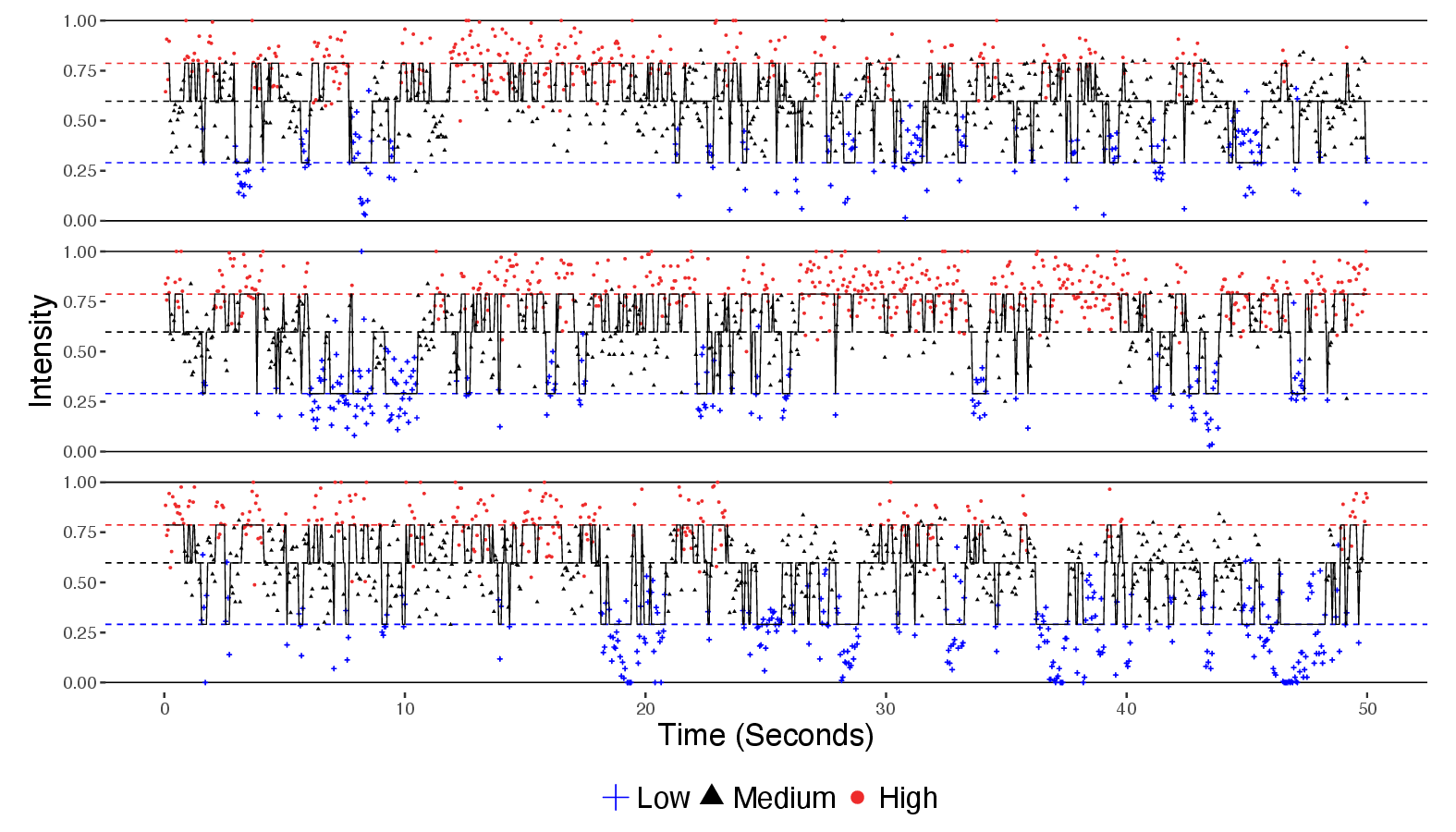}
  \caption{Quantum dot analysis: Intensity series of 3 randomly selected quantum dots in Cluster 1. }
  \label{fig:Group1}
\end{figure}

\begin{figure}[H]
  \centering
  \includegraphics[width = 0.9\linewidth]{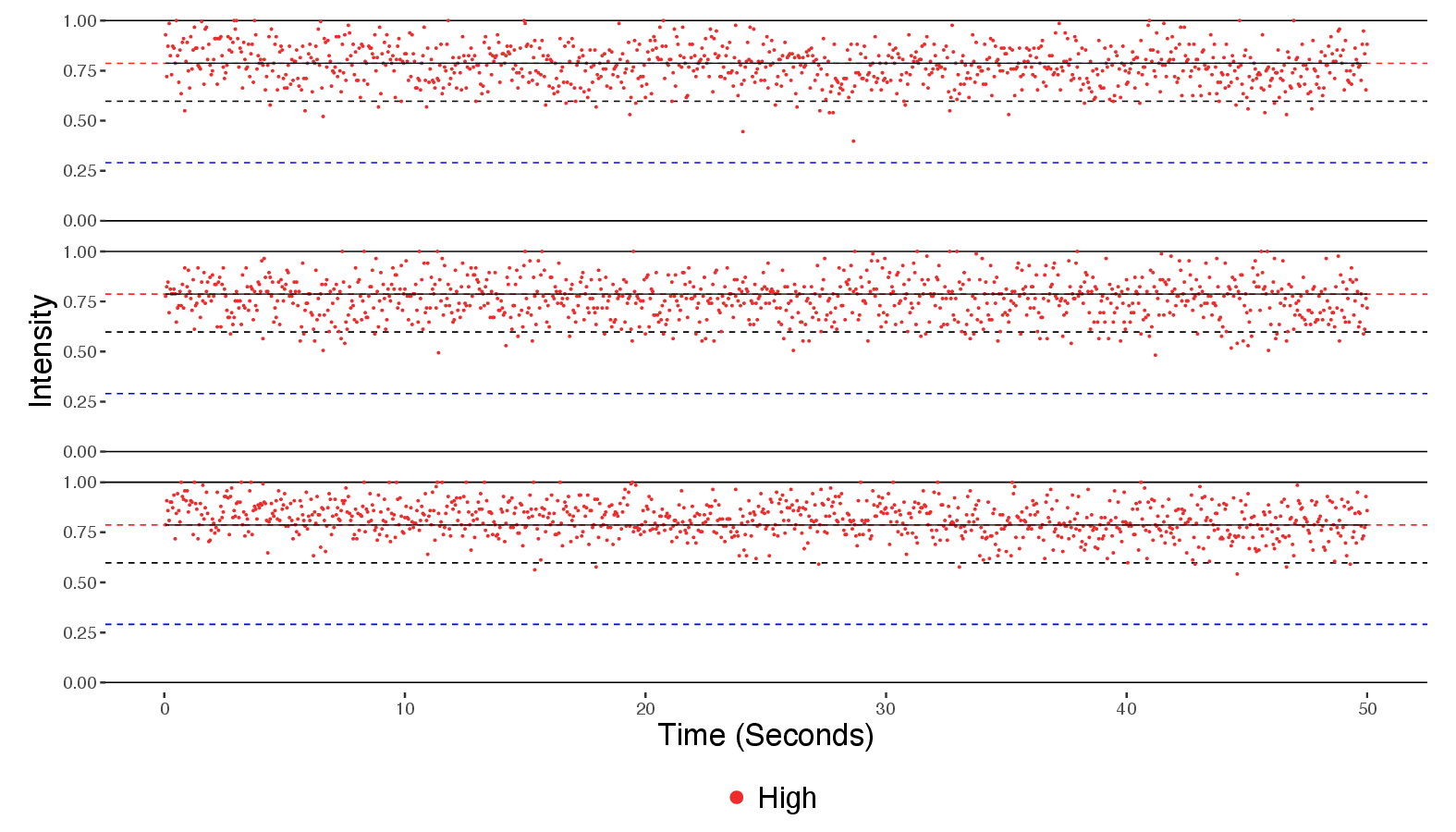}
  \caption{Quantum dot analysis: Intensity series of 3 randomly selected quantum dots in Cluster 2.}
  \label{fig:Group2}
\end{figure}

\begin{figure}[H]
  \centering
  \includegraphics[width = 0.9\linewidth]{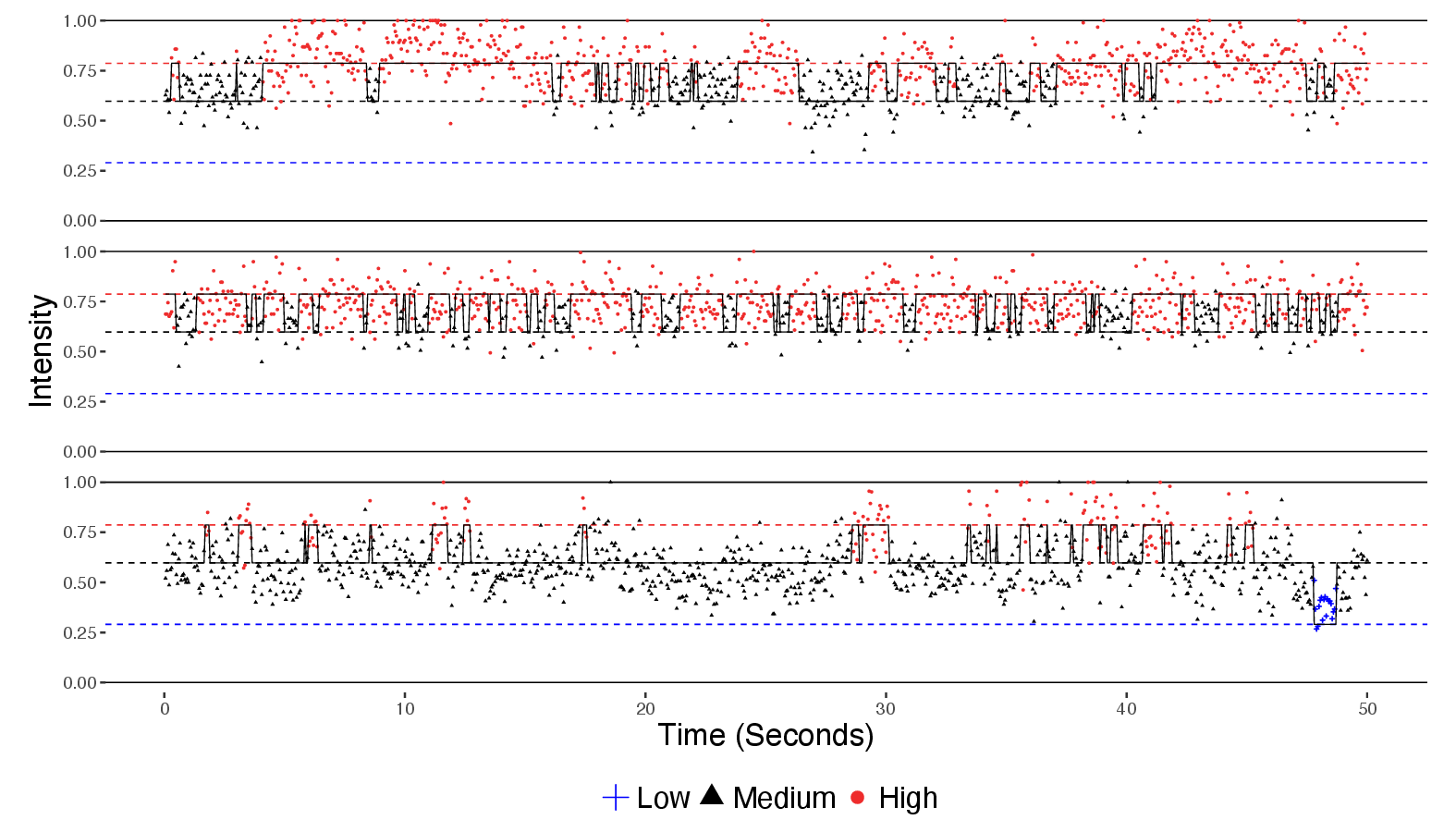}
  \caption{Quantum dot analysis: Intensity series of 3 randomly selected quantum dots in Cluster 3.}
  \label{fig:Group3}
\end{figure}

The estimated transition matrices for the three clusters are as follows:
\begin{equation}
  \widehat{\boldsymbol{\Pi}}_1 =
  \begin{bmatrix}
  0.848 & 0.127 & 0.025 \\
  0.060 & 0.794 & 0.146 \\
  0.017 & 0.186 & 0.796
  \end{bmatrix}
  , \widehat{\boldsymbol{\Pi}}_2 = 
  \begin{bmatrix}
  0.795 & 0.205 & 0.000 \\
  0.000 & 0.994 & 0.006 \\
  0.000 & 0.001 & 0.999
  \end{bmatrix}
  , \widehat{\boldsymbol{\Pi}}_3 = 
  \begin{bmatrix}
  0.938 & 0.057 & 0.005 \\
  0.013 & 0.924 & 0.063 \\
  0.001 & 0.067 & 0.932
  \end{bmatrix}.
  \nonumber
\end{equation}
The corresponding stationary distributions from these estimated transition matrices are as follows:
\begin{equation}
    \widehat{\boldsymbol{\pi}}_1 = 
    \begin{bmatrix}
    0.213 & 0.443 & 0.344
    \end{bmatrix} 
    , \widehat{\boldsymbol{\pi}}_2 = 
    \begin{bmatrix}
    0.000 & 0.143 & 0.857
    \end{bmatrix} 
    , \widehat{\boldsymbol{\pi}}_3 = 
    \begin{bmatrix}
    0.104 & 0.461 & 0.435
    \end{bmatrix}. 
    \nonumber
\end{equation}
Based on the estimated MHMM-$\beta$ model and according to the Bayes rule, there are 28, 54, and 46 QDs in Clusters 1, 2, and 3, respectively. To visualize, we randomly pick 3 dots from each cluster and presented their intensity series along with their estimated state series in Figures \ref{fig:Group1}, \ref{fig:Group2}, and \ref{fig:Group3}, respectively.

In Cluster 1, the QDs exhibit about 80\%-85\% chance of remaining at the same state and display relatively more frequent transitions between states comparing to QDs in the other two clusters. Specifically, the most frequent transitions happen from low to medium, medium to high, and high to medium, and the chances are 12.7\%, 14.6\%, and 18.6\%, respectively. From the stationary distribution, these QDs spend relatively even amount of time in the three states, i.e., 21.3\%, 44.3\%, and 34.4\%, respectively. These behaviors are consistent with the examples visualized in Figure \ref{fig:Group1}. 

In Cluster 2, the QDs exhibit extremely high probabilities of staying in the medium or high-intensity states, and more importantly, the probabilities of transiting into the low-intensity state are essentially zero. This results in a stationary distribution showing that these QDs spend most of their time in the high-intensity state with probability 91.1\%, a much smaller amount of time in the medium-intensity state with probability 8.9\%, and ultimately no time at all in the low-intensity state. In Figure \ref{fig:Group2}, the three randomly selected QDs from this cluster all stayed in the high-intensity state throughout the continuous excitation.

In Cluster 3, the QDs exhibit relatively high chance of remaining at the same state (92.2\% - 93.8\%) and display relatively less frequent transitions between states, comparing to QDs in Cluster 1. The most frequent transitions remains the same as in Cluster 1, i.e., from low to medium, medium to high, and high to medium, but the chances are reduced to 5.7\%, 6.3\%, and 6.7\%, respectively. Consequently, the stationary distribution reveals that these QDs spend about 45\% of time in either mediam and high-intensity states and only 10\% of the time in the low-intensity state. These are reflected in Figure \ref{fig:Group3}, in which two dots only spent time in median and high-intensity states while the third dot also briefly visited the low-intensity state.

\subsection{Implications}

The analysis of quantum dot data using the proposed MHMM-$\beta$ model revealed critical insights into the intensity fluctuation behaviors and clustering patterns of QDs under continuous excitation. Specifically, three distinct clusters were identified, each characterized by unique transition dynamics and stationary distributions of intensity states. Cluster 1 exhibited relatively balanced transitions across all intensity states, with QDs spending comparable time in low, medium, and high-intensity states. Cluster 2 represented highly stable QDs predominantly residing in the high-intensity state, exhibiting minimal transitions and no time in the low-intensity state. Cluster 3 showcased QDs with moderate transition dynamics, spending most of their time in medium and high-intensity states and only briefly visiting the low-intensity state. These findings align with the hypothesized ``blinking'' and ``flickering'' styles of intensity fluctuations and provide a rigorous statistical characterization of QD behaviors, offering valuable insights for their applications in chemistry and materials science.

\section{Conclusion \& Discussion}
\label{sec:6}

This study introduces a novel statistical approach for analyzing the intensity fluctuation patterns of colloidal quantum dots (QDs), a critical problem in the fields of chemistry and materials science. The main challenge lies in the inherent complexity of the data: the intensity measurements exhibit intricate patterns influenced by unobserved states and are subject to substantial variation across individual QDs. Through a close collaboration with chemistry researchers, our proposed analytical pipeline and the mixture hidden Markov model with an inflated Beta distribution (MHMM-$\beta$) address these challenges by simultaneously modeling the intensity fluctuations and clustering the QDs based on their transition dynamics. This integrative approach captures the shared structure among QDs while allowing for individual differences, providing a more comprehensive understanding of their individual and collective behaviors under continuous excitation.

The findings from our analysis highlight the method's effectiveness and its ability to generate interpretable results that align with principles of chemistry and physics. By identifying distinct clusters of QDs, each characterized by unique transition dynamics and stationary distributions, our approach provides chemists with valuable insights into the interplay between QD intensity states and their fluctuation styles.

There are several promising directions for future research. One critical and promising task is the joint analysis of photon lifetime and intensity. Recent chemical studies have emphasized the importance of understanding the relationship between photon lifetime, which reflects aspects of the QD microenvironment, and intensity fluctuations, which capture macroenvironmental behaviors. Developing methods that incorporate photon lifetime into the integrative analysis would enable us to reveal dependencies and connections between these two dimensions, offering a more holistic view of QD properties. Further methodological advancements could focus on developing more efficient algorithms for computation and implementing more accurate methods for model selection. 
From a broader perspective, the proposed approach could be adapted to address other scientific problems involving complex temporal data with state-dependent dynamics. For example, similar methods could be applied in neuroimaging to study brain activity patterns or in ecology to analyze animal movement behaviors.  

\section*{Acknowledgment}
Zhao and Chen were partially supported by NSF Grant CHE-2203854.


\appendix

\clearpage
\appendix
\noindent {\bf \LARGE Supplementary Material}
\section{Details on the EM Algorithm}\label{supp:comp}

\subsection{Derivation of the likelihood functions}\label{sec:appendix:likelihood}

The complete-data log-likelihood function is given by:
\begin{equation}
\begin{split}
l(\boldsymbol{\Theta};\boldsymbol{X},\boldsymbol{S},\boldsymbol{Z}) 
= & \sum_{i=1}^N \log \{ \prod_{k=1}^K [\delta_{k}\boldsymbol{\pi}_{k,\boldsymbol{s}_{i,1}} f(x_{i,1} \mid \boldsymbol{s}_{i,1}) \prod_{t=2}^T (\boldsymbol{\Pi}_k)_{\boldsymbol{s}_{i,t-1},\boldsymbol{s}_{i,t}} f(x_{i,t} \mid \boldsymbol{s}_{i,t}) ]^{I(z_{i,k}=1)} \} \\
= & \sum_{i=1}^N \sum_{k=1}^K I(z_{i,k}=1) \log(\delta_{k})
+ \sum_{i=1}^N \sum_{k=1}^K I(z_{i,k}=1) \sum_{p = 1}^M I(s_{i,1,p} = 1)\log(\boldsymbol{\pi}_{k,p}) \\
& + \sum_{i=1}^N \sum_{k=1}^K I(z_{i,k}=1) \log(\prod_{t=2}^T (\boldsymbol{\Pi}_k)_{\boldsymbol{s}_{i,t-1},\boldsymbol{s}_{i,t}}) \\
& + \sum_{i=1}^N \sum_{k=1}^K I(z_{i,k}=1) \log( \prod_{t=1}^T f(x_{i,t} \mid \boldsymbol{s}_{i,t}))  \\
= & \sum_{i=1}^N \sum_{k=1}^K I(z_{i,k}=1) \log(\delta_{k})
+ \sum_{i=1}^N \sum_{k=1}^K I(z_{i,k}=1) \sum_{p = 1}^M I(s_{i,1,p} = 1)\log(\boldsymbol{\pi}_{k,p}) \\
& + \sum_{i=1}^N \sum_{k=1}^K I(z_{i,k}=1) \{\sum_{t=2}^T\sum_{p=1}^M\sum_{q=1}^M I(s_{i,t-1,p} = 1,s_{i,t,q} = 1) \log ((\boldsymbol{\Pi}_k)_{p,q})\}  \\
& + \sum_{i=1}^N \sum_{k=1}^K I(z_{i,k}=1) \{ \sum_{t=1}^T \sum_{p=1}^M I(s_{i,t,p} = 1) \log f(x_{i,t} \mid s_{i,t,p} = 1) \}.
\end{split}\notag
\end{equation}

\subsection{E-Step}\label{sec:appendix:estep}

Let $\boldsymbol{\Theta}^{(j)}$ denote the set of parameter estimates from $j^{th}$ iteration. The expected complete-data log-likelihood function is then given by:
\begin{eqnarray}
Q(\boldsymbol{\Theta} \mid \boldsymbol{\Theta}^{(j)}) & = & E_{(\boldsymbol{S},\boldsymbol{Z}) \sim p(\cdot,\cdot \mid \boldsymbol{X},\boldsymbol{\Theta}^{(j)}) }[l(\boldsymbol{\Theta};\boldsymbol{X},\boldsymbol{S},\boldsymbol{Z}) \mid \boldsymbol{X}, \boldsymbol{\Theta}^{(j)} ] \nonumber \\
& = & E_{\boldsymbol{S}\sim p(\cdot \mid \boldsymbol{X},\boldsymbol{Z},\boldsymbol{\Theta}^{(j)}) }\{E_{\boldsymbol{Z} \sim p(\cdot \mid \boldsymbol{X},\boldsymbol{\Theta}^{(j)})}[ l(\boldsymbol{\Theta};\boldsymbol{X},\boldsymbol{S},\boldsymbol{Z})]\}.
\label{fl:AppendEstep}
\end{eqnarray}
In \eqref{fl:AppendEstep}, we rewrite the integration w.r.t. the joint distribution $(\boldsymbol{S},\boldsymbol{Z})$ as a sequence of expectations, which could simplify the computation. 

\noindent\textbf{(i) Compute $I(z_{i,k}=1)$: }
\begin{eqnarray}
E[I(z_{i,k}=1) \mid \boldsymbol{x}_i, \boldsymbol{\Theta}^{(j)}] & = & P(z_{i,k} = 1 \mid \boldsymbol{x}_i, \boldsymbol{\Theta}^{(j)}) \nonumber \\
& = & \frac{P(z_{i,k} = 1, \boldsymbol{x}_i \mid \boldsymbol{\Theta}^{(j)})}{\sum_{k' = 1,\dots,K} P(z_{i,k'} = 1, \boldsymbol{x}_i\mid\boldsymbol{\Theta}^{(j)})}.
\label{fl:EstepZ}
\end{eqnarray}
In \eqref{fl:EstepZ}, the key part is $P(z_{i,k} = 1, \boldsymbol{x}_i \mid \boldsymbol{\Theta}^{(j)})$. To calculate this probability, we adopt the idea of forward algorithm in dynamic programming \citep{Rabiner1989_IEEEproc}. That is, the forward probabilities are defined as 
\begin{equation}
  \alpha_{i,t,k,h} = P((x_{i,1},\dots,x_{i,t})\trans,s_{i,t,h} = 1 \mid z_{i,k} = 1, \boldsymbol{\Theta}^{(j)}).
  \notag
\end{equation}
 The value of $\boldsymbol{\alpha}_{i,t,k}$ can be determined through iterative computation as follows.
\begin{eqnarray}
\boldsymbol{\alpha}_{i,t,k} & = & \boldsymbol{\pi}_{k} \circ (f(x_{i,t})\mid s_{i,t,h} = 1,\boldsymbol{\Theta}^{(j)})_{h=1,\dots,M}\trans,\textnormal{when }t = 1,
\nonumber \\
\boldsymbol{\alpha}_{i,t,k} & = & (\boldsymbol{\Pi}_k\trans \cdot \boldsymbol{\alpha}_{i,t-1,k}) \circ (f(x_{i,t})\mid s_{i,t,h} = 1,\boldsymbol{\Theta}^{(j)})_{h=1,\dots,M}\trans, \textnormal{when } t > 1,
\notag
\end{eqnarray}
where $\circ$ is the Hadamard product. 

Using the forward probabilities at the last time point $T$, we have that $ P(z_{i,k} = 1, \boldsymbol{x}_i \mid \boldsymbol{\Theta}^{(j)}) = \lVert \boldsymbol{\alpha}_{i,T,k} \rVert_1 = \sum_{h=1}^M \alpha_{i,T,k,h}$ and 
\begin{eqnarray}
\tau_{i,k}(\boldsymbol{\Theta}^{(j)}) \equiv E[I(z_{i,k}=1)\mid \boldsymbol{x}_i,\boldsymbol{\Theta}^{(j)}] & = & \frac{ \lVert \boldsymbol{\alpha}_{i,T,k} \rVert_1}{\sum_{k' = 1,\dots,K} \lVert \boldsymbol{\alpha}_{i,T,k'} \rVert_1} \nonumber \\
& = & \frac{\sum_{h=1}^M \alpha_{i,T,k,h}}{\sum_{k' = 1,\dots,K} \sum_{h=1}^M \alpha_{i,T,k',h}}.
\label{fl:EstepZfinished}
\end{eqnarray}
Here, following \citet{Rabiner1989_IEEEproc}, we have denoted $E[I(z_{i,k}=1)\mid\boldsymbol{x}_i,\boldsymbol{\Theta}^{(j)}]$ as $\tau_{i,k}(\boldsymbol{\Theta}^{(j)})$.\\

After evaluating the inner expectation, the outer expectation is computed as follows:

\noindent\textbf{(ii) Compute $I(s_{i,t,h}=1)$ :}

Only $E[I(s_{i,t,h}=1)]$ is explicitly solved and $E[I(s_{i,1,h}=1)]$ is treated as a special case of the former.

\begin{eqnarray}
E[I(s_{i,t,h} = 1)\mid \boldsymbol{z}_i,\boldsymbol{x}_i,\boldsymbol{\Theta}^{(j)}]
& = & P(s_{i,t,h} = 1\mid \boldsymbol{z}_i,\boldsymbol{x}_i,\boldsymbol{\Theta}^{(j)})\nonumber \\
& = & \frac{P(s_{i,t,h} = 1, \boldsymbol{x}_i\mid \boldsymbol{z}_i,\boldsymbol{\Theta}^{(j)})}{\sum_{h' = 1,\dots,M} P(s_{i,t,h'} = 1 ,\boldsymbol{x}_i\mid \boldsymbol{z}_i,\boldsymbol{\Theta}^{(j)})}.
\label{fl:EstepSZ}
\end{eqnarray}
In \eqref{fl:EstepSZ}, the key part is $P(s_{i,t,h} = 1, \boldsymbol{x}_i \mid \boldsymbol{z}_i,\boldsymbol{\Theta}^{(j)})$. To calculate this probability, we adopt the idea of backward algorithm in dynamic programming \citep{Rabiner1989_IEEEproc}. We write $\boldsymbol{\beta}_{i,t,k} = (\beta_{i,t,k,h})_{h=1,\dots,M}\trans$, where
\begin{equation}
\beta_{i,t,k,h} = P((x_{i,t+1},\dots,x_{i,T})\trans , s_{i,t,h} = 1 \mid z_{i,k} = 1,\boldsymbol{\Theta}^{(j)}).
\notag
\end{equation}
The value of $\boldsymbol{\beta}_{i,t,k}$ can be determined by iterative calculations as follows.
\begin{eqnarray}
\boldsymbol{\beta}_{i,t,k} & = & \boldsymbol{1}, \textnormal{when } t = T, \nonumber \\
\boldsymbol{\beta}_{i,t,k} & = & \boldsymbol{\Pi}_k\trans \cdot [(f(x_{i,t})\mid s_{i,t,h} = 1,\boldsymbol{\Theta}^{(j)})_{h=1,\dots,M} \circ \boldsymbol{\beta}_{i,t+1,k}], \textnormal{when } t < T.
\notag 
\end{eqnarray}
Combine the forward probabilities and backward probabilities, \eqref{fl:EstepSZ} can be updated as:

\begin{eqnarray}
\gamma_{i,t,k,h}(\boldsymbol{\Theta}^{(j)}) 
& \equiv & E[I(s_{i,t,h}=1) \mid z_{i,k}=1, \boldsymbol{x}_i, \boldsymbol{\Theta}^{(j)}] = P(s_{i,t,h}=1 \mid z_{i,k}=1,\boldsymbol{x}_i, \boldsymbol{\Theta}^{(j)})\nonumber \\
& = & \frac{\boldsymbol{\alpha}_{i,t,k} \circ \boldsymbol{\beta}_{i,t,k}}{\boldsymbol{\alpha}_{i,t,k}\trans \cdot \boldsymbol{\beta}_{i,t,k}}, \nonumber
\label{fl:EstepSZfinished}
\end{eqnarray}
and
\begin{eqnarray}
\eta_{i,t,h}(\boldsymbol{\Theta}^{(j)}) &\equiv&
E[I(s_{i,t,h}=1)\mid \boldsymbol{x}_i,\boldsymbol{\Theta}^{(j)}] =  \sum_{k=1}^K P(s_{i,t,h} = 1 \mid z_{i,k} = 1, \boldsymbol{x}_i, \boldsymbol{\Theta}^{(j)}) P(z_{i,k} = 1 \mid \boldsymbol{x}_i,\boldsymbol{\Theta}^{(j)}) \nonumber \\
& = &\sum_{k=1}^K\gamma_{i,t,k,h}(\boldsymbol{\Theta}^{(j)})\tau_{i,k}(\boldsymbol{\Theta}^{(j)}). \nonumber
\end{eqnarray} 

We use $\gamma_{i,t,k,h}(\boldsymbol{\Theta}^{(j)})$ to denote $ E[I(s_{i,t,h}=1)\mid z_{i,k} = 1,\boldsymbol{x}_i,\boldsymbol{\Theta}^{(j)}]$ and use $\eta_{i,t,h}(\boldsymbol{\Theta}^{(j)})$ to denote $ E[I(s_{i,t,h}=1)\mid \boldsymbol{x}_i,\boldsymbol{\Theta}^{(j)}]$. \\

\noindent\textbf{(iii) Compute $I(s_{i,t-1,p} = 1,s_{i,t,q} = 1)$:}
\begin{eqnarray}
\xi_{i,t,k,p,q}(\boldsymbol{\Theta}^{(j)}) 
& = & E[I(s_{i,t-1,p} = 1,s_{i,t,q} = 1) \mid \boldsymbol{z}_i, \boldsymbol{x}_i, \boldsymbol{\Theta}^{(j)}]\nonumber\\
& = & P[s_{i,t-1,p} = 1,s_{i,t,q} = 1\mid\boldsymbol{z}_i,\boldsymbol{x}_i,\boldsymbol{\Theta}^{(j)}] \nonumber \\
& = & \frac{\alpha_{i,t-1,k,p} (\boldsymbol{\Pi}_k)_{p,q} f(x_{i,t}\mid s_{i,t,q} = 1) \beta_{i,t,k,q}}{\sum_{p' = 1}^M\sum_{q' = 1}^M \alpha_{i,t-1,k,p'} (\boldsymbol{\Pi}_k)_{p',q'} f(x_{i,t}\mid s_{i,t,q'} = 1) \beta_{i,t,k,q'}}.
\notag
\end{eqnarray}
We denote $E[I(s_{i,t-1,p}=1, s_{i,t,q} = 1)\mid z_{i,k} = 1,\boldsymbol{x}_i,\boldsymbol{\Theta}^{(j)}]$ as $\xi_{i,t,k,p,q}(\boldsymbol{\Theta}^{(j)})$.

\subsection{M-Step}
In the M-step, the parameters are updated by maximizing the expected complete-data log-likelihood $ Q(\boldsymbol{\Theta}\mid \boldsymbol{\Theta}^{(j)}) $, which has been reformulated in terms of the intermediate quantities $\tau_{i,k}(\boldsymbol{\Theta}^{(j)})$, $\gamma_{i,t,k,h}(\boldsymbol{\Theta}^{(j)})$, $\eta_{i,t,h}(\boldsymbol{\Theta}^{(j)})$ and $\xi_{i,t,k,p,q}(\boldsymbol{\Theta}^{(j)})$.

\begin{eqnarray}
Q(\boldsymbol{\Theta}\mid \boldsymbol{\Theta}^{(j)})
    & = & \sum_{i=1}^N \sum_{k=1}^K \tau_{i,k}(\boldsymbol{\Theta}^{(j)}) \log(\delta_{k}) + \sum_{i=1}^N \sum_{k=1}^K \tau_{i,k}(\boldsymbol{\Theta}^{(j)}) \log( \boldsymbol{\pi}_{k,\boldsymbol{s}_{i,1}})\notag\\
    & & + \sum_{i=1}^N \sum_{k=1}^K \tau_{i,k}(\boldsymbol{\Theta}^{(j)}) \{\sum_{t=2}^T\sum_{p=1}^M\sum_{q=1}^M \xi_{i,t,k,p,q}(\boldsymbol{\Theta}^{(j)}) \log (\boldsymbol{\Pi}_k)_{p,q}\}  \nonumber \\
    & & + \sum_{i=1}^N \sum_{k=1}^K \tau_{i,k}(\boldsymbol{\Theta}^{(j)}) \{ \sum_{t=1}^T \sum_{p=1}^M \gamma_{i,t,k,p}(\boldsymbol{\Theta}^{(j)}) \log f(x_{i,t} \mid s_{i,t,p} = 1) \}.  \nonumber
\notag 
\end{eqnarray}

\noindent\textbf{(i) Maximize with respect to $\delta_k$:}

\begin{eqnarray}
\hat{\boldsymbol{\delta}}^{(j+1)} & = & \argmax_{\boldsymbol{\delta}} \sum_{i=1}^N \sum_{k=1}^K \tau_{i,k}(\boldsymbol{\Theta}^{(j)}) \log( \delta_{k}) , \qquad s.t. \sum_{k=1}^K \delta_k = 1.
\label{fl:MstepDelta}
\end{eqnarray}
Applying the method of Lagrange multipliers to maximize \eqref{fl:MstepDelta} leads to the following derivation:
\begin{eqnarray}
\mathcal{L} & = & \sum_{i=1}^N \sum_{k=1}^K \tau_{i,k}(\boldsymbol{\Theta}^{(j)}) \log( \delta_{k}) + \lambda(\sum_{k=1}^K \delta_k - 1), \nonumber\\
\nabla(\mathcal{L}) & = & (
\frac{ \sum_{i=1}^N \tau_{i,1}(\boldsymbol{\Theta}^{(j)})}{\delta_1} - \lambda, \dots,
\frac{ \sum_{i=1}^N \tau_{i,K}(\boldsymbol{\Theta}^{(j)})}{\delta_K} - \lambda,
\sum_{k=1}^K \delta_k  - 1)\trans, \nonumber \\ 
& \textnormal{Set } & \nabla(\mathcal{L}) = \boldsymbol{0}, \nonumber \\
& \textnormal{Solution:} & \lambda =  \sum_{i=1}^N \sum_{k'=1}^K \tau_{i,k'}(\boldsymbol{\Theta}^{(j)}) = N, \nonumber \\
\delta_k^{(j+1)} & = & \frac{\sum_{i=1}^N \tau_{i,k}(\boldsymbol{\Theta}^{(j)})}{N}.
\notag
\end{eqnarray}

\noindent\textbf{(ii) Maximization with respect to $\boldsymbol{\pi}_{k,\boldsymbol{s}_{i,1}}$:}

\begin{eqnarray}
\widehat{\boldsymbol{\pi}}^{(j+1)} & = & \argmax_{\boldsymbol{\pi}} \sum_{i=1}^N \sum_{k=1}^K \tau_{i,k}(\boldsymbol{\Theta}^{(j)}) \log( \hat{\boldsymbol{\pi}}^{(j)} ), \nonumber \\
 & s.t.& \sum_{h=1}^M \boldsymbol{\pi}_{k,h} = 1, \forall k \in \{1,\dots,K\}. 
 \nonumber
\label{fl:MstepPi}
\end{eqnarray}

Using the method of Lagrange multipliers, the result is  
\begin{equation}
\widehat{\boldsymbol{\pi}}_{k,h}^{(j+1)} = \frac{\sum_{i=1}^N \tau_{i,k}(\boldsymbol{\Theta}^{(j)})\gamma_{i,1,k,h}(\boldsymbol{\Theta}^{(j)})}{\sum_{i=1}^N \tau_{i,k}(\boldsymbol{\Theta}^{(j)})}.\notag
\end{equation}

\noindent\textbf{(iii) Maximization with respect to $(\boldsymbol{\Pi}_k)_{p,q}$:}

\begin{eqnarray}
\widehat{\boldsymbol{\Pi}}_k^{(j+1)} & = & \argmax_{\boldsymbol{\Pi}_k}\sum_{i=1}^N \sum_{k=1}^K \tau_{i,k}(\boldsymbol{\Theta}^{(j)}) \{\sum_{t=2}^T\sum_{p=1}^M\sum_{q=1}^M \xi_{i,t,k,p,q}(\boldsymbol{\Theta}^{(j)}) \log (\boldsymbol{\Pi}_k)_{p,q} \}, \nonumber \\ 
& s.t. & \sum_{q=1}^M (\boldsymbol{\Pi}_k)_{p,q} = 1, 
\qquad \forall   p \in \{1,\dots,M\}, k \in \{1,\dots,K\}. \nonumber
\label{fl:MstepAk}
\end{eqnarray}

Using the method of Lagrange multipliers, the result is 
\begin{equation}
(\boldsymbol{\Pi}_k)_{p,q}^{(j+1)} = \frac{ \sum_{i=1}^N \tau_{i,k}(\boldsymbol{\Theta}^{(j)}) \sum_{t=2}^T \xi_{i,t,k,p,q}(\boldsymbol{\Theta}^{(j)}) }{\sum_{i=1}^N \tau_{i,k}(\boldsymbol{\Theta}^{(j)}) \sum_{t=2}^T \sum_{q'=1}^M \xi_{i,t,k,p,q'}(\boldsymbol{\Theta}^{(j)})
}.\notag
\end{equation} 

\noindent\textbf{(iv) Maximization with respect to  $(a_h,b_h,\epsilon_h^{(0)},\epsilon_h^{(1)})_{h = 1,\dots,M}$:}

\begin{equation}
\begin{aligned}
(\hat{a}_h,\hat{b}_h,\hat{\epsilon}_h^{(0)},\hat{\epsilon}_h^{(1)})
= & \argmax_{(a_h,b_h,\epsilon_h^{(0)},\epsilon_h^{(1)})} \sum_{i=1}^N \sum_{k=1}^K \tau_{i,k}(\boldsymbol{\Theta}^{(j)}) \{ \sum_{t=1}^T  \gamma_{i,t,k,h}(\boldsymbol{\Theta}^{(j)}) \log f(x_{i,t} \mid s_{i,t,h} = 1) \} \\
= & \argmax_{(a_h,b_h,\epsilon_h^{(0)},\epsilon_h^{(1)})} \sum_{i=1}^N \sum_{k=1}^K \tau_{i,k}(\boldsymbol{\Theta}^{(j)}) \{ \sum_{t=1}^T \gamma_{i,t,k,h}(\boldsymbol{\Theta}^{(j)}) \\
& + [I(x_{i,t} = 0)\log(\epsilon_h^{(0)}) + I(x_{i,t} = 1)\log(\epsilon_h^{(1)})  \\
& + I(0 < x_{i,t} < 1)\log (1 - \epsilon_h^{(0)} - \epsilon_h^{(1)}) + I(0 < x_{i,t} < 1) \log f(x_{i,t}\mid a_h,b_h) ] \}.
\end{aligned}
\label{fl:MstepBeta}
\end{equation}

The maximization problem in \eqref{fl:MstepBeta} can be decomposed into two components, $\epsilon_h^{(0)},\epsilon_h^{(1)}$ and the $a_h, b_h$, which can be updated separately.\\

\noindent\textbf{(v) Update $(\epsilon_h^{(0)},\epsilon_h^{(1)})_{h = 1,\dots,M}$: }
\begin{equation}
\begin{split}
(\hat{\epsilon}_h^{(0)},\hat{\epsilon}_h^{(1)} = 
& \argmax_{(\epsilon_h^{(0)}
,\epsilon_h^{(1)})} \sum_{i=1}^N \sum_{k=1}^K \tau_{i,k}(\boldsymbol{\Theta}^{(j)}) \{ \sum_{t=1}^T \gamma_{i,t,k,h}(\boldsymbol{\Theta}^{(j)})\\
&[
I(x_{i,t} = 0)\log(\epsilon_h^{(0)}) +
I(x_{i,t} = 1)\log(\epsilon_h^{(1)}) +
I(0 < x_{i,t} < 1)\log( 1 - \epsilon_h^{(0)} - \epsilon_h^{(1)})  ] \}.
\end{split}
\label{fl:Mstepepsilon}
\end{equation}
The solutions to \eqref{fl:Mstepepsilon} are:
\begin{eqnarray}
(\epsilon_h^{(0)})^{(j+1)} & = & \frac{ \sum_{i=1}^N \sum_{k=1}^K \tau_{i,k}(\boldsymbol{\Theta}^{(j)}) \sum_{t=1}^T \gamma_{i,t,k,h}(\boldsymbol{\Theta}^{(j)}) I(x_{i,t} = 0)}{ \sum_{i=1}^N \sum_{k=1}^K \tau_{i,k}(\boldsymbol{\Theta}^{(j)}) \sum_{t=1}^T \gamma_{i,t,k,h}(\boldsymbol{\Theta}^{(j)})}, \nonumber \\
(\epsilon_h^{(1)})^{(j+1)} & = & \frac{ \sum_{i=1}^N \sum_{k=1}^K \tau_{i,k}(\boldsymbol{\Theta}^{(j)}) \sum_{t=1}^T \gamma_{i,t,k,h}(\boldsymbol{\Theta}^{(j)}) I(x_{i,t} = 1)}{ \sum_{i=1}^N \sum_{k=1}^K \tau_{i,k}(\boldsymbol{\Theta}^{(j)}) \sum_{t=1}^T \gamma_{i,t,k,h}(\boldsymbol{\Theta}^{(j)})}.
\nonumber
\label{fl:updateepsilon}
\end{eqnarray}

\noindent\textbf{(vi) Update $(a_h,b_h)_{h = 1,\dots,M}$:}

\begin{equation}
(\hat{a}_h,\hat{b}_h) = \argmax_{(a_h, b_h)} \sum_{i=1}^N \sum_{k=1}^K \tau_{i,k}(\boldsymbol{\Theta}^{(j)}) \{ \sum_{t=1}^T \gamma_{i,t,k,h}(\boldsymbol{\Theta}^{(j)}) I(0 < x_{i,t} < 1) \log(f(x_{i,t}\mid a_h,b_h)) \}.
\label{fl:Mstepab}
\end{equation}
We employ the Newton-Raphson method to solve \eqref{fl:Mstepab}. Let $(a_h^{(r)},b_h^{(r)})\trans$ denote the parameter estimates obtained at $r^{\mbox{th}}$ iteration. The gradient vector is defined as $\boldsymbol{g} =(g_1,g_2)\trans$, where $g_1$ and $g_2$ denote the first-order partial derivatives of the objective function in \eqref{fl:Mstepab} with respect to $a_h^{(r)}$ and $b_h^{(r)}$, respectively. The Hessian matrix $\boldsymbol{G}$ is defined as $\boldsymbol{G} = \begin{pmatrix}   
 g_{11} & g_{12} \\       
 g_{21} & g_{22}          
\end{pmatrix} $, where each $g_{ij}$ denotes the second-order partial derivative with respect to the corresponding parameters. To simplify the notation in the subsequent derivations, we denote the objective function in \eqref{fl:Mstepab} by $\mathcal{L}$ and let the $\psi$ denote the Digamma function, i.e., $\psi(x) = \frac{\partial}{\partial x} \ln{\Gamma(x)}$.
\begin{eqnarray*}
g_1 & = &  \frac{\partial \mathcal{L} }{\partial a_h}|_{(a_h,b_h)=(a_h^{(r)},b_h^{(r)})}  \\
    & = & \sum_{i=1}^N \sum_{k=1}^K \tau_{i,k}(\boldsymbol{\Theta}^{(j)}) \{ \sum_{t=1}^T \gamma_{i,t,k,h}(\boldsymbol{\Theta}^{(j)}) I(0 < x_{i,t} < 1) [ \psi(a_h^{(r)}+b_h^{(r)}) - \psi(a_h^{(r)}) + \log(x_{i,t})] \} ,  \\
g_2 & =& \frac{\partial \mathcal{L} }{\partial b_h}|_{(a_h,b_h)=(a_h^{(r)},b_h^{(r)})} \\
    & = & \sum_{i=1}^N \sum_{k=1}^K \tau_{i,k}(\boldsymbol{\Theta}^{(j)}) \{ \sum_{t=1}^T \gamma_{i,t,k,h}(\boldsymbol{\Theta}^{(j)}) I(0 < x_{i,t} < 1) [\psi(a_h^{(r)}+b_h^{(r)}) - \psi(b_h^{(r)}) + \log(1-x_{i,t})] \},\\
g_{11} & = & \frac{\partial g_1}{\partial a_h}|_{(a_h,b_h)=(a_h^{(r)},b_h^{(r)})} \\
       & = & \sum_{i=1}^N \sum_{k=1}^K \tau_{i,k}(\boldsymbol{\Theta}^{(j)}) \{ \sum_{t=1}^T \gamma_{i,t,k,h}(\boldsymbol{\Theta}^{(j)}) I(0 < x_{i,t} < 1) [\psi'(a_h^{(r)} + b_h^{(r)}) - \psi'(a_h^{(r)})], \\
g_{12} & = & \frac{\partial g_1}{\partial b_h}|_{(a_h,b_h)=(a_h^{(r)},b_h^{(r)})} \\
       & = &\sum_{i=1}^N \sum_{k=1}^K \tau_{i,k}(\boldsymbol{\Theta}^{(j)}) \{ \sum_{t=1}^T \gamma_{i,t,k,h}(\boldsymbol{\Theta}^{(j)}) I(0 < x_{i,t} < 1) [\psi'(a_h^{(r)} + b_h^{(r)}) ],  \\
g_{21} & = & \frac{\partial g_2}{\partial a_h}|_{(a_h,b_h)=(a_h^{(r)},b_h^{(r)})}  = g_{12} , \\
g_{22} & = & \frac{\partial g_2}{\partial b_h}|_{(a_h,b_h)=(a_h^{(r)},b_h^{(r)})}  \\
       & = & \sum_{i=1}^N \sum_{k=1}^K \tau_{i,k}(\boldsymbol{\Theta}^{(j)}) \{ \sum_{t=1}^T \gamma_{i,t,k,h}(\boldsymbol{\Theta}^{(j)}) I(0 < x_{i,t} < 1) [\psi'(a_h^{(r)} + b_h^{(r)}) - \psi'(b_h^{(r)})]. 
\end{eqnarray*}
New estimates are given as follows: 
\begin{eqnarray}
(a_h^{(r+1)}, b_h^{(r+1)})\trans & = & (a_h^{(r)}, b_h^{(r)})\trans - \boldsymbol{G}^{-1}\boldsymbol{g} .\nonumber
\end{eqnarray}
\subsection{Pseudo-Code} 

The algorithms are summarized as follows.
\begin{algorithm}[H]
  \caption{Newton-Raphson Method for $(a_h,b_h)$}
  \label{alg:Newton}
  \begin{algorithmic}
    \State \textbf{Input:} $a_h^{(j)}$, $b_h^{(j)}$, $\tau$, $\gamma$
    \State Control parameters: $\epsilon_{\mbox{NRtol}} = 1e-5, \mbox{maxit}_{\mbox{NR}} = 1e+3$
    \State Initial values:  $a^{(0)} \gets a_h^{(j)}, b^{(0)} \gets b_h^{(j)}$
    \State Initialize the temporary parameters: $\Delta \gets 1, r \gets 0$
    
    \While{$r < \mbox{maxit}_{\mbox{NR}}$ \mbox{and} $\Delta \geq \epsilon_{\mbox{NRtol}}$}
    \State Calculate $g_1,g_2,g_{11},g_{12},g_{21},g_{22}$ 
    \State Write $\boldsymbol{g}$ and $\boldsymbol{G}$
    \State $(a^{(r+1)}, b^{(r+1)})\trans = (a^{(r)}, b^{(r)})\trans - \boldsymbol{G}^{-1}\boldsymbol{g}$
    \State $\Delta \gets \frac{\lVert (a^{(r+1)}-a^{(r)}, b^{(r+1)} - b^{(r)}) \rVert_2^2}{\lVert (a^{(r)}, b^{(r)}) \rVert_2^2}$
    \State $r \gets r + 1$
    \EndWhile
    \State $( a_h^{(j+1)} \gets a^{(r+1)}, b_h^{(j+1)} \gets b^{(j+1)})$
    \State \textbf{Output:} $ (a_h^{(j+1)}, b_h^{(j+1)})$
  \end{algorithmic}
\end{algorithm}

\begin{algorithm}[H]
  \caption{Core EM-algorithm}
  \label{alg:EMalgorithm}
  \begin{algorithmic}
    \State \textbf{Input:} $\boldsymbol{X}$, $K$, $M$, $N$
    \State Set up control parameters: $\epsilon_{\mbox{EM-tol}} = 1e-5$, $maxit_{\mbox{EM}} = 1e+3$
    \State Set up initial values: $\boldsymbol{\Theta}^{(0)}$
    \State Initialize the temporary parameters: $j \gets 0$
    \Repeat \Comment{EM-algorithm}
    \State  \Comment{E-step}
    \For {$i = 1,\dots, N$}  
    \State Using current parameters $\boldsymbol{\Theta}^{(j)}$
    \State Calculate forward-backward probabilities, $\boldsymbol{\alpha}_{i,t,k},\boldsymbol{\beta}_{i,t,k}$
    \State Calculate $\tau_{i,k}(\boldsymbol{\Theta}^{(j)})$, $\gamma_{i,t,k,h}(\boldsymbol{\Theta}^{(j)})$, $\xi_{i,t,k,p,q}(\boldsymbol{\Theta}^{(j)})$
    \EndFor

    \State \Comment{M-step}
    \For {$k = 1,\dots, K$}
    \State Update $\delta_k^{(j+1)}$, $\boldsymbol{\pi}_k$
    \For { $\forall (p,q) \in (1,\dots,M)\times (1,\dots,M)$} 
    \State Update $(\boldsymbol{\Pi}_k)_{p,q}^{(j+1)}$
    \EndFor
    \EndFor

    \For {$h = {1,\dots,M}$} 
    \State Update $(\epsilon_h^{(0)})^{(j+1)} $ and $(\epsilon_h^{(1)})^{(j+1)} $ 
    \State Update $a_h^{(j+1)}, b_h^{(j+1)}$ using Algorithm \ref{alg:Newton}
    \EndFor
    \State $\boldsymbol{\Theta}^{(j+1)}$ is the collection of all new parameters from M-step
    
    \Until{Convergence of log-likelihood or max iteration reached}

    \State \textbf{Output}: Estimated hidden states, estimated cluster and $\widehat{\boldsymbol{\Theta}}$, the MLE of parameters.    
  \end{algorithmic}
\end{algorithm}

\section{Inference}\label{supp:inference}
\subsection{Variance Estimation}\label{sec:appendix:variance}

To describe the uncertainty of 0/1 inflated beta parameters, we utilize the SEM algorithm mentioned in \citet{Meng1991}. 
Among all parameters in $\boldsymbol{\Theta}$, we use  $\boldsymbol{\Phi}$ to denote the subset of parameters associated with the 0/1 inflated beta distribution. Let $\boldsymbol{\mathcal{F}}$ be a mapping from $\boldsymbol{\Phi}$ to $\boldsymbol{\Phi}$, such that $\boldsymbol{\Phi}^{(j+1)} = \boldsymbol{\mathcal{F}}(\boldsymbol{\Phi}^{(j)})$. Moreover, the variance-covariance matrix of the parameters, denoted as $\boldsymbol{V}$, is defined and its value is given as follows:
\begin{equation}
   \begin{split}
   \boldsymbol{V} & = \boldsymbol{I}^{-1}_{oc} + \boldsymbol{U}, \\
   \boldsymbol{U} & = \boldsymbol{I}^{-1}_{oc} \frac{\partial \boldsymbol{\mathcal{F}(\Phi)}}{\partial \boldsymbol{\Phi}}|_{\boldsymbol{\Phi} = \widehat{\boldsymbol{\Phi}}} (\boldsymbol{I} - \frac{\partial \boldsymbol{\mathcal{F}(\Phi)}}{\partial \boldsymbol{\Phi}}|_{\boldsymbol{\Phi} = \widehat{\boldsymbol{\Phi}}})^{-1},
   \end{split}
   \label{eq:SEM} 
\end{equation}
where $\boldsymbol{I}^{-1}_{oc}$ is the inverse of the observed data information matrix, and $\frac{\partial \boldsymbol{\mathcal{F}(\Phi)}}{\partial \boldsymbol{\Phi}}|_{\boldsymbol{\Phi} = \widehat{\boldsymbol{\Phi}}}$ is the derivative of $\boldsymbol{\mathcal{F}}$ evaluated at $\boldsymbol{\Phi} = \widehat{\boldsymbol{\Phi}}$.

As the first key part in \eqref{eq:SEM}, $\frac{\partial \boldsymbol{\mathcal{F}(\Phi)}}{\partial \boldsymbol{\Phi}}|_{\boldsymbol{\Phi} = \widehat{\boldsymbol{\Phi}}}$ is calculated as follows. For simplicity, let $m_{p,q}$ denote the $(p,q)$-th element of the $\frac{\partial \boldsymbol{\mathcal{F}(\Phi)}}{\partial \boldsymbol{\Phi}}|_{\boldsymbol{\Phi} = \widehat{\boldsymbol{\Phi}}}$. To facilitate the presentation of details in SEM algorithm, we write $\widehat{\boldsymbol{\Phi}}$ as a vector of length $4M$, $(\widehat{\epsilon}_h^{(0)}, \widehat{\epsilon}_h^{(1)},\widehat{a}_h,\widehat{b}_h)\trans_{h = 1,\ldots,M}$. And then, we define the one-step-rollback estimation on $p$-th parameter in the $\widehat{\boldsymbol{\Phi}}$ as $\widehat{\boldsymbol{\Phi}}^{(j)}_{p}$, that is, replace the $p$-th parameter in MLE by its estimate in $j$-th iteration. For example, if $\widehat{a}_1$ is the $p$-th element in $\widehat{\boldsymbol{\Phi}}$,  $\widehat{\boldsymbol{\Phi}}^{(j)}_{p}$ can be written as $ (\widehat{\epsilon}_1^{(0)}, \widehat{\epsilon}_1^{(1)},a_h^{(j)},\widehat{b}_h,\ldots,\widehat{\epsilon}_M^{(0)}, \widehat{\epsilon}_M^{(1)},\widehat{a}_M,\widehat{b}_M)\trans$.

\begin{algorithm}[H]
\caption{The SEM Algorithm for Variance-covariance Matrix}
\label{alg:SEM}
\begin{algorithmic}
\State \textbf{Input:} The MLE of parameters, $\widehat{\boldsymbol{\Phi}}$, and the estimated parameters in $j$-th iteration, $\boldsymbol{\Phi}^{(j)}$.
\For{$p = 1$  to $|\widehat{\boldsymbol{\Phi}}|$}
    \State Write $\widehat{\boldsymbol{\Phi}}^{(j)}_{p}$, the one-step-rollback estimation on the $p$-th parameter in the $\widehat{\boldsymbol{\Phi}}$.
    
    \State Treat $\widehat{\boldsymbol{\Phi}}^{(j)}_{p}$ as current estimation, and run one iteration of EM to obtain $\widetilde{\boldsymbol{\Phi}}^{(j+1)}_{p}$.
    
    \State The $p$-th row of $\frac{\partial \boldsymbol{\mathcal{F}(\Phi)}}{\partial \boldsymbol{\Phi}}|_{\boldsymbol{\Phi} = \widehat{\boldsymbol{\Phi}}}$ can be calculated as  $\frac{\widetilde{\boldsymbol{\Phi}}^{(j+1)}_{p} - \widehat{\boldsymbol{\Phi}}}{\boldsymbol{\Phi}^{(j)}(p) - \widehat{\boldsymbol{\Phi}}(p)}$, where $\widehat{\boldsymbol{\Phi}}(p)$ stands for the $p$-th element in $\widehat{\boldsymbol{\Phi}}$ and similar for $\boldsymbol{\Phi}^{(j)}$.
\EndFor
\State Repeat this algorithm on all iteration $j$, until values in $\frac{\partial \boldsymbol{\mathcal{F}(\Phi)}}{\partial \boldsymbol{\Phi}}|_{\boldsymbol{\Phi} = \widehat{\boldsymbol{\Phi}}}$ become stable.
\State \textbf{Output:} $\frac{\partial \boldsymbol{\mathcal{F}(\Phi)}}{\partial \boldsymbol{\Phi}}|_{\boldsymbol{\Phi} = \widehat{\boldsymbol{\Phi}}}$.
\end{algorithmic}
\end{algorithm}

As the second key part in \eqref{eq:SEM}, $\boldsymbol{I}^{-1}_{oc}$ is the inverse of the observed data information matrix evaluated at MLE. Similar to the situation discussed in \citet{Meng1991}, considering the direct computation of the observed-data information matrix is very difficult, we firstly calculate the complete-data information matrix $\boldsymbol{I}_{c}$ and then take its expectation over the conditional distribution $f(\boldsymbol{S},\boldsymbol{Z} \mid \boldsymbol{X}, \boldsymbol{\Phi})$ evaluated at $\boldsymbol{\Phi} = \widehat{\boldsymbol{\Phi}}$. 
The complete data information matrix $\boldsymbol{I}_{c}$ is calculated by $\boldsymbol{I}_{c}(\boldsymbol{\Theta} \mid \boldsymbol{X,S,Z}) = -\frac{\partial^2 \log f(\boldsymbol{X,S,Z} \mid \boldsymbol{\Theta})}{\partial \boldsymbol{\Theta}^2}$, which fully depends on the complete data likelihood.
And $\boldsymbol{I}_{oc} = E[\boldsymbol{I}_{c}(\boldsymbol{\Theta} \mid \boldsymbol{X},\boldsymbol{S},\boldsymbol{Z})\mid \boldsymbol{X}, \boldsymbol{\Phi} ]|_{\boldsymbol{\Phi} = \widehat{\boldsymbol{\Phi}}}$. 
That is, in our case, $\boldsymbol{\Phi} = (\epsilon_1^{(0)}, \epsilon_1^{(1)}, a_1, b_1, \ldots, \epsilon_M^{(0)}, \epsilon_M^{(1)}, a_M, b_M)$ and $\boldsymbol{I}_{oc}$ can be represented as a sparse matrix,
\[
\begin{bmatrix}
\boldsymbol{I}_{oc}^{(1)} & \boldsymbol{0} & \boldsymbol{0} & \boldsymbol{0} \\
\boldsymbol{0} & \boldsymbol{I}_{oc}^{(2)} & \boldsymbol{0} & \boldsymbol{0} \\
\boldsymbol{0} & \boldsymbol{0} & \cdots & \boldsymbol{0} \\
\boldsymbol{0} & \boldsymbol{0} & \boldsymbol{0} & \boldsymbol{I}_{oc}^{(M)}
\end{bmatrix}
\]

Combining two key parts could give the estimate of the variance-covariance matrix, $\boldsymbol{V}$.

\section{Additional Simulation Results}\label{supp:sim}
\subsection{Additional Results from Simulation}

We consider a third scenario that is slightly modified from Scenario 1, to make the three clusters slightly more distinct. The transition matrices are given by 
\begin{equation}
\boldsymbol{\Pi}_1=
\begin{bmatrix} 
0.848 & 0.127 & 0.025 \\ 
0.099 & 0.735 & 0.166 \\ 
0.017 & 0.186 & 0.796 
\end{bmatrix}
,
\boldsymbol{\Pi}_2=
\begin{bmatrix} 
0.795 & 0.205 & 0.000 \\
0.000 & 0.994 & 0.006 \\ 
0.000 & 0.001 & 0.999 
\end{bmatrix}
,
\boldsymbol{\Pi}_3=
\begin{bmatrix}
0.938 & 0.057 & 0.005 \\ 
0.013 & 0.924 & 0.063 \\ 
0.001 & 0.067 & 0.932 
\end{bmatrix},
\nonumber
\end{equation}
and the parameters for the three states are set as 
\begin{align*}
(a_1,b_1,\epsilon_1^{(0)},\epsilon_1^{(1)}) & = (2.000, 5.000, 0.025, 0.000),\\ (a_2,b_2,\epsilon_2^{(0)},\epsilon_2^{(1)}) & = (10.000, 7.000, 0.000, 0.001),\\ 
(a_3,b_3,\epsilon_3^{(0)},\epsilon_3^{(1)}) & = (12.000, 3.000, 0.000, 0.020).
\end{align*}

The simulation results are showing in Tables \ref{table:scenario3:balanced} and \ref{table:scenario3:unbalanced}. The observations from these results are all consistent with those reported in Section 4 
of the main manuscript; we thus omit the details here. 

\begin{table}[H]
  \centering
    \scriptsize
\caption{Simulation: Results for Scenario 3 with balanced partition. For better presentation, values in $\mbox{Er}(\widehat{\boldsymbol{\delta}})$ and $\mbox{Er}(\widehat{\boldsymbol{\theta}})$ are multiplied by $10$, and values in $\mbox{Er}(\widehat{\mu})$ and $\mbox{Er}(\widehat{\sigma}^2)$ are multiplied by $10^2$.}
\label{table:scenario3:balanced}
\makebox[\textwidth][c]{
\begin{tabular}{ l c | c c c | c c c}
  Method & Length ($T$) &  $\mbox{Er}(\widehat{\mu})$ & $\mbox{Er}(\widehat{\sigma}^2)$ & $\mbox{Er}(\widehat{\boldsymbol{\delta}})$ & $\mbox{Er}(\widehat{\boldsymbol{\theta}})$ & $\mbox{Er}(\widehat{\boldsymbol{\Pi}})$ & \mbox{CC} \\
  \hline
  \multicolumn{8}{c}{Balanced, Sample size ($N$) = 100}\\
  \hline
MHMM-$\beta^*$ &  \multirow{5}{*}{250} & 0.36 (0.02) & 0.66 (0.04) & 0.00 (0.00) & 2.96 (0.12) & 0.12 (0.01) &  \\ 
  HMM-G &   & 14.50 (0.12) & 10.15 (0.07) & 3.42 (0.08) &  & 2.18 (0.03) & 0.54 (0.00) \\ 
  HMM-$\beta$ &   & 21.77 (0.18) & 9.23 (0.09) & 3.93 (0.06) &  & 2.21 (0.03) & 0.53 (0.00) \\ 
  MHMM-G &   & 0.94 (0.04) & 2.94 (0.08) & 0.89 (0.12) &  & 0.60 (0.07) & 0.93 (0.01) \\ 
  MHMM-$\beta$ &   & 0.37 (0.02) & 0.66 (0.04) & 0.46 (0.10) & 3.06 (0.12) & 0.30 (0.05) & 0.97 (0.01) \\  
  \hline
MHMM-$\beta^*$ & \multirow{5}{*}{500}  & 0.25 (0.01) & 0.51 (0.03) & 0.00 (0.00) & 2.15 (0.09) & 0.10 (0.01) &  \\ 
  HMM-G &   & 14.00 (0.09) & 8.68 (0.05) & 3.75 (0.05) &  & 2.14 (0.02) & 0.51 (0.00) \\ 
  HMM-$\beta$ &   & 21.07 (0.13) & 6.24 (0.06) & 3.76 (0.05) &  & 1.98 (0.02) & 0.52 (0.00) \\ 
  MHMM-G &   & 0.92 (0.03) & 2.90 (0.07) & 0.93 (0.14) &  & 0.56 (0.06) & 0.93 (0.01) \\ 
  MHMM-$\beta$ &   & 0.26 (0.01) & 0.52 (0.03) & 0.45 (0.11) & 2.29 (0.09) & 0.31 (0.05) & 0.97 (0.01) \\  
  \hline
MHMM-$\beta^*$ &  \multirow{5}{*}{1000} & 0.18 (0.01) & 0.36 (0.02) & 0.00 (0.00) & 1.45 (0.05) & 0.10 (0.01) &  \\ 
  HMM-G &   & 13.07 (0.06) & 7.75 (0.03) & 3.56 (0.04) &  & 2.01 (0.02) & 0.58 (0.01) \\ 
  HMM-$\beta$ &   & 19.63 (0.11) & 4.62 (0.06) & 3.87 (0.02) &  & 1.95 (0.01) & 0.58 (0.00) \\ 
  MHMM-G &   & 0.85 (0.02) & 2.79 (0.05) & 0.71 (0.13) &  & 0.42 (0.05) & 0.95 (0.01) \\ 
  MHMM-$\beta$ &   & 0.21 (0.01) & 0.41 (0.02) & 0.62 (0.13) & 1.61 (0.07) & 0.38 (0.06) & 0.96 (0.01) \\ 
  \hline
MHMM-$\beta^*$ & \multirow{5}{*}{2000} & 0.13 (0.01) & 0.26 (0.01) & 0.00 (0.00) & 1.11 (0.04) & 0.09 (0.01) &  \\ 
  HMM-G &   & 12.23 (0.05) & 7.44 (0.02) & 2.87 (0.14) &  & 1.77 (0.04) & 0.70 (0.02) \\ 
  HMM-$\beta$ &   & 18.49 (0.08) & 3.81 (0.03) & 3.41 (0.13) &  & 1.84 (0.03) & 0.67 (0.01) \\ 
  MHMM-G &   & 0.92 (0.02) & 2.90 (0.03) & 0.64 (0.12) &  & 0.36 (0.04) & 0.95 (0.01) \\ 
  MHMM-$\beta$ &   & 0.16 (0.01) & 0.32 (0.02) & 0.30 (0.10) & 1.42 (0.07) & 0.26 (0.04) & 0.98 (0.01) \\  
  \hline     
  \end{tabular} 
}
\end{table}

\begin{table}[H]
  \centering
    \scriptsize
\caption{Simulation: Results for Scenario 3 with unbalanced partition. For better presentation, values in $\mbox{Er}(\widehat{\boldsymbol{\delta}})$ and $\mbox{Er}(\widehat{\boldsymbol{\theta}})$ are multiplied by $10$, and values in $\mbox{Er}(\widehat{\mu})$ and $\mbox{Er}(\widehat{\sigma}^2)$ are multiplied by $10^2$.}
\label{table:scenario3:unbalanced}
\makebox[\textwidth][c]{
\begin{tabular}{ l c | c c c | c c c}
  Method & Length ($T$) &  $\mbox{Er}(\widehat{\mu})$ & $\mbox{Er}(\widehat{\sigma}^2)$ & $\mbox{Er}(\widehat{\boldsymbol{\delta}})$ & $\mbox{Er}(\widehat{\boldsymbol{\theta}})$ & $\mbox{Er}(\widehat{\boldsymbol{\Pi}})$ & \mbox{CC} \\
\hline  
  \multicolumn{8}{c}{Unbalanced, Sample size ($N$) = 100}\\
  \hline
MHMM-$\beta^*$ & \multirow{5}{*}{250} & 0.34 (0.02) & 0.60 (0.04) & 0.00 (0.00) & 2.97 (0.14) & 0.15 (0.01) &  \\ 
  HMM-G &   & 7.42 (0.08) & 8.99 (0.05) & 1.38 (0.05) &  & 2.26 (0.02) & 0.79 (0.01) \\ 
  HMM-$\beta$ &   & 14.89 (0.18) & 6.16 (0.08) & 1.43 (0.03) &  & 2.34 (0.02) & 0.81 (0.00) \\ 
  MHMM-G &   & 1.76 (0.04) & 4.97 (0.08) & 0.94 (0.08) &  & 0.89 (0.07) & 0.91 (0.01) \\ 
  MHMM-$\beta$ &   & 0.35 (0.02) & 0.63 (0.04) & 0.85 (0.11) & 3.08 (0.15) & 0.45 (0.05) & 0.93 (0.01) \\ 
  \hline
MHMM-$\beta^*$ & \multirow{5}{*}{500} & 0.24 (0.01) & 0.47 (0.02) & 0.00 (0.00) & 2.20 (0.10) & 0.12 (0.01) &  \\ 
  HMM-G &   & 7.72 (0.07) & 8.19 (0.05) & 1.35 (0.03) &  & 2.24 (0.02) & 0.82 (0.00) \\ 
  HMM-$\beta$ &   & 14.80 (0.14) & 4.15 (0.06) & 1.49 (0.03) &  & 2.27 (0.02) & 0.81 (0.00) \\ 
  MHMM-G &   & 1.68 (0.03) & 4.89 (0.06) & 0.55 (0.08) &  & 0.72 (0.07) & 0.95 (0.01) \\ 
  MHMM-$\beta$ &   & 0.26 (0.01) & 0.50 (0.03) & 0.24 (0.07) & 2.33 (0.11) & 0.36 (0.05) & 0.98 (0.01) \\ 
\hline
  MHMM-$\beta^*$ & \multirow{5}{*}{1000} & 0.18 (0.01) & 0.35 (0.02) & 0.00 (0.00) & 1.75 (0.08) & 0.13 (0.01) &  \\ 
  HMM-G &   & 7.68 (0.05) & 7.79 (0.03) & 1.26 (0.02) &  & 2.21 (0.02) & 0.84 (0.00) \\ 
  HMM-$\beta$ &   & 14.08 (0.12) & 3.07 (0.05) & 1.43 (0.02) &  & 2.22 (0.02) & 0.83 (0.00) \\ 
  MHMM-G &   & 1.63 (0.02) & 4.81 (0.04) & 0.32 (0.08) &  & 0.50 (0.05) & 0.97 (0.01) \\ 
  MHMM-$\beta$ &   & 0.20 (0.01) & 0.39 (0.02) & 0.09 (0.04) & 2.05 (0.10) & 0.24 (0.04) & 0.99 (0.00) \\ 
  \hline 
MHMM-$\beta^*$ & \multirow{5}{*}{2000} & 0.13 (0.01) & 0.23 (0.01) & 0.00 (0.00) & 1.21 (0.05) & 0.11 (0.01) &  \\ 
  HMM-G &   & 7.49 (0.04) & 7.49 (0.02) & 1.10 (0.04) &  & 2.07 (0.04) & 0.87 (0.01) \\ 
  HMM-$\beta$ &   & 13.40 (0.08) & 2.65 (0.03) & 1.31 (0.02) &  & 2.18 (0.02) & 0.85 (0.00) \\ 
  MHMM-G &   & 1.62 (0.02) & 4.81 (0.03) & 0.09 (0.05) &  & 0.29 (0.03) & 0.99 (0.00) \\ 
  MHMM-$\beta$ &   & 0.15 (0.01) & 0.28 (0.01) & 0.05 (0.03) & 1.46 (0.07) & 0.17 (0.02) & 1.00 (0.00) \\ 
  \hline
  \end{tabular}
}  
\end{table}

\subsection{Simulation Study on Model Selection}
Determining the true number of clusters is crucial. In practice, we advocate the use of both information criteria and domain knowledge. Here, we perform a simulation study to evaluate the performance of different information criteria. 

We consider widely used information criteria such as the Akaike Information Criterion (AIC) \citep{Akaike1974_IEEE} and the Bayesian Information Criterion (BIC) \citep{Schwarz1978}. 
\begin{equation}
\begin{split}
\mbox{AIC} &= -2 l(\widehat{\boldsymbol{\Theta}} \mid \boldsymbol{X}) + 2|\widehat{\boldsymbol{\Theta}}|, \\
\mbox{BIC} &= -2 l(\widehat{\boldsymbol{\Theta}} \mid \boldsymbol{X}) + \log(NT) |\widehat{\boldsymbol{\Theta}}|.
\label{eq:AICandBIC}
\nonumber
\end{split}
\end{equation}
Here, $N$ represents the total number of QDs, $T$ denotes the length of each QD, $l(\widehat{\boldsymbol{\Theta}} \mid \boldsymbol{X})$ is the observed data log-likelihood, and $|\widehat{\boldsymbol{\Theta}}|$ represents the total number of free parameters in the model.  

In addition to AIC and BIC, we also consider the Integrated Completed Likelihood (ICL) criterion \citep{ICL}, which is commonly used in cluster analysis.
\begin{equation}
\begin{split}
\mbox{ICL} & = -2 l(\widehat{\boldsymbol{\Theta}}\mid \boldsymbol{X},\widehat{\boldsymbol{Z}},\widehat{\boldsymbol{S}}) + \log(NT) |\widehat{\boldsymbol{\Theta}}|.
\label{eq:ICLdefinition}
\nonumber
\end{split}
\end{equation} 
Here, the ICL is an approximation to the complete data likelihood, $l(\widehat{\boldsymbol{\Theta}} \mid \boldsymbol{X},\boldsymbol{Z},\boldsymbol{S})$.

We simulate data under Scenario 3 described above. We then fit the simulated data using MHMM-$\beta$ with varying numbers of clusters and states. For each setting, the model selection is conducted based on AIC, BIC, and ICL.

Table \ref{table:merged_findk} reports the frequency with which each number of clusters is selected by ICL, AIC, and BIC when fitting the MHMM-$\beta$ model over 100 replications. In almost all settings, the correct choice of three clusters dominates, confirming that each criterion can recover the true number of clusters most of the time. Nonetheless, the extent of agreement varies. In balanced scenarios, all three criteria many occasionally over-select the number of clusters, where ICL and BIC favor exactly three clusters more often than AIC. In unbalanced partitions, selecting the correct three-cluster model becomes slightly more challenging; both AIC and BIC more frequently suggest additional clusters, while ICL may also under-select the number of clusters.  

These results illustrate the well-known differences in how each criterion penalizes model complexity and cluster separation; while they often converge on three clusters, discrepancies sometimes arise in cases of overlapping or unevenly sized clusters. 

\begin{table}[H]
\centering
\caption{Proportion (\%) of times each number of clusters is selected by ICL, AIC, and BIC under Scenario 3. Rows correspond to different partitions (balanced or unbalanced), sample sizes, and series lengths.}
\label{table:merged_findk}
\begin{tabular}{l c | r r r r r | r r r r r | r r r r r}
\hline
\multirow{2}{*}{N} & \multirow{2}{*}{T} 
& \multicolumn{5}{c|}{ICL} 
& \multicolumn{5}{c|}{AIC} 
& \multicolumn{5}{c}{BIC}\\
 & 
 & \#1 & \#2 & \#3 & \#4 & \#5 
 & \#1 & \#2 & \#3 & \#4 & \#5
 & \#1 & \#2 & \#3 & \#4 & \#5 \\
\hline
\multicolumn{16}{c}{\textbf{Balanced Partition}} \\
\hline
\multirow{2}{*}{50} & 1000 
& 0 & 0 & 81 & 18 & 1 
& 0 & 0 & 66 & 33 & 1
& 0 & 0 & 83 & 17 & 0 \\
  & 2000 
& 0 & 0 & 74 & 25 & 1
& 0 & 0 & 66 & 29 & 5
& 0 & 0 & 77 & 22 & 1 \\
\multirow{2}{*}{100} & 1000 
& 0 & 0 & 74 & 23 & 3 
& 0 & 0 & 69 & 28 & 3
& 0 & 0 & 79 & 21 & 0 \\
  & 2000 
& 0 & 0 & 77 & 20 & 3
& 0 & 0 & 74 & 21 & 5
& 0 & 0 & 85 & 15 & 0 \\
\hline
\multicolumn{16}{c}{\textbf{Unbalanced Partition}} \\
\hline
\multirow{2}{*}{50} & 1000
& 0 & 12 & 63 & 23 & 2
& 0 & 0 & 59 & 35 & 6
& 0 & 0 & 66 & 29 & 5 \\
 & 2000
& 0 & 0 & 79 & 20 & 1
& 0 & 0 & 71 & 25 & 4
& 0 & 0 & 80 & 20 & 0 \\
\multirow{2}{*}{100} & 1000
& 0 & 2 & 67 & 24 & 7
& 0 & 0 & 63 & 29 & 8
& 0 & 0 & 68 & 27 & 5 \\
 & 2000
& 0 & 0 & 78 & 18 & 4
& 0 & 0 & 75 & 22 & 3
& 0 & 0 & 85 & 15 & 0 \\
\hline
\end{tabular}
\end{table}

\section{Additional Results in Data Analysis}\label{supp:app}
\subsection{Results from 3-Cluster Model}
\begin{equation}
  \widehat{\boldsymbol{\Pi}}_1 =
  \begin{bmatrix}
  0.848 & 0.127 & 0.025 \\
  0.060 & 0.794 & 0.146 \\
  0.017 & 0.186 & 0.796
  \end{bmatrix}
  , \widehat{\boldsymbol{\Pi}}_2 = 
  \begin{bmatrix}
  0.795 & 0.205 & 0.000 \\
  0.000 & 0.994 & 0.006 \\
  0.000 & 0.001 & 0.999
  \end{bmatrix}
  , \widehat{\boldsymbol{\Pi}}_3 = 
  \begin{bmatrix}
  0.938 & 0.057 & 0.005 \\
  0.013 & 0.924 & 0.063 \\
  0.001 & 0.067 & 0.932
  \end{bmatrix}.
  \nonumber
\end{equation}
\begin{equation}
    \hat{\boldsymbol{\pi}}_1 = 
    \begin{bmatrix}
    0.213 & 0.443 & 0.344
    \end{bmatrix} 
    , \hat{\boldsymbol{\pi}}_2 = 
    \begin{bmatrix}
    0.000 & 0.089 & 0.911
    \end{bmatrix} 
    , \hat{\boldsymbol{\pi}}_3 = 
    \begin{bmatrix}
    0.106 & 0.459 & 0.435
    \end{bmatrix}. 
    \nonumber
\end{equation}
The numbers of QDs in the three clusters are as follows.
\begin{table}[h!]
\centering
\begin{tabular}{|c|c|c|c|}
\hline
Cluster & 1 & 2 & 3 \\ \hline
Count & 28 & 54 & 46 \\ \hline
\end{tabular}
\end{table}

\subsection{Results from 4-Cluster Model}
\begin{equation}
  \begin{split}
    \widehat{\boldsymbol{\Pi}}_1 &=
    \begin{bmatrix}
       0.842 & 0.133 & 0.025 \\
       0.067 & 0.784 & 0.149 \\
       0.022 & 0.212 & 0.766
    \end{bmatrix}
    , \quad \widehat{\boldsymbol{\Pi}}_2 =
    \begin{bmatrix}
       0.799 & 0.201 & 0.000 \\
       0.000 & 0.993 & 0.007 \\
       0.000 & 0.001 & 0.999
    \end{bmatrix}, \\
    \widehat{\boldsymbol{\Pi}}_3 &=
    \begin{bmatrix}
       0.899 & 0.085 & 0.016 \\
       0.017 & 0.842 & 0.141 \\
       0.002 & 0.082 & 0.917
    \end{bmatrix}
    , \quad \widehat{\boldsymbol{\Pi}}_4 =
    \begin{bmatrix}
       0.946 & 0.053 & 0.001 \\
       0.014 & 0.941 & 0.045 \\
       0.001 & 0.094 & 0.905
    \end{bmatrix}.
  \end{split}
  \nonumber
\end{equation}

\begin{equation}
  \begin{split}
    \hat{\boldsymbol{\pi}}_1 &=
    \begin{bmatrix}
    0.235 & 0.452 & 0.313
    \end{bmatrix}
    , \quad \hat{\boldsymbol{\pi}}_2 =
    \begin{bmatrix}
    0.000 & 0.125 & 0.875
    \end{bmatrix}, \\
    \hat{\boldsymbol{\pi}}_3 &=
    \begin{bmatrix}
    0.069 & 0.343 & 0.588
    \end{bmatrix}
    , \quad \hat{\boldsymbol{\pi}}_4 =
    \begin{bmatrix}
    0.154 & 0.573 & 0.273
    \end{bmatrix}.
  \end{split}
  \nonumber
\end{equation}

The numbers of QDs in the four clusters are as follows.
\begin{table}[h!]
\centering
\begin{tabular}{|c|c|c|c|c|}
\hline
Cluster & 1 & 2 & 3 & 4 \\ \hline
Count & 23 & 53 & 27 & 25 \\ \hline
\end{tabular}
\end{table}

To analyze the relationship between the 3-cluster and 4-cluster solutions, we present the correspondence of cluster assignments as follows. 
\begin{table}[ht]
\centering
\begin{tabular}{rrrr}
  \hline
 & 1 & 2 & 3 \\ 
  \hline
1 &  23 &   0 &   0 \\ 
  2 &   0 &  53 &   0 \\ 
  3 &   5 &   1 &  21 \\ 
  4 &   0 &   0 &  25 \\ 
   \hline
\end{tabular}
\end{table}

In the 4-cluster model, Clusters 1 and 2 closely align with Clusters 1 and 2 from the 3-cluster model, with similar estimated transition matrices and equilibrium distributions. Clusters 3 and 4 in the 4-cluster model primarily result from a split of Cluster 3 in the 3-cluster model. Specifically, the new Cluster 3 is associated with more time spent in the high state and exhibits a higher probability of transitioning to and remaining in that state, as indicated by the last column of $\widehat{\boldsymbol{\Pi}}_3$. In contrast, Cluster 4 shows longer durations in the medium state and a stronger tendency to remain there compared to Cluster 3.

\subsection{Results from 5-Cluster Model}
\begin{equation}
  \begin{aligned}
    &\widehat{\boldsymbol{\Pi}}_1 =
    \begin{bmatrix}
    0.814 & 0.160 & 0.026 \\
    0.073 & 0.774 & 0.153 \\
    0.029 & 0.280 & 0.692
    \end{bmatrix},
    \quad \widehat{\boldsymbol{\Pi}}_2 =
    \begin{bmatrix}
    0.000 & 1.000 & 0.000 \\
    0.000 & 0.993 & 0.007 \\
    0.000 & 0.000 & 1.000
    \end{bmatrix},
    \quad \widehat{\boldsymbol{\Pi}}_3 =
    \begin{bmatrix}
    0.889 & 0.092 & 0.019 \\
    0.036 & 0.785 & 0.179 \\
    0.006 & 0.132 & 0.863
    \end{bmatrix}, \\[10pt]
    &\widehat{\boldsymbol{\Pi}}_4 =
    \begin{bmatrix}
    0.936 & 0.063 & 0.001 \\
    0.024 & 0.922 & 0.055 \\
    0.002 & 0.154 & 0.844
    \end{bmatrix},
    \quad \widehat{\boldsymbol{\Pi}}_5 =
    \begin{bmatrix}
    0.929 & 0.051 & 0.021 \\
    0.003 & 0.944 & 0.053 \\
    0.001 & 0.041 & 0.958
    \end{bmatrix}.
  \end{aligned}
  \nonumber
\end{equation}

\begin{equation}
  \begin{split}
    & \hat{\boldsymbol{\pi}}_1 =
    \begin{bmatrix}
    0.237 & 0.497 & 0.266
    \end{bmatrix}
    , \quad \hat{\boldsymbol{\pi}}_2 =
    \begin{bmatrix}
     0.000 & 0.000 & 1.000
    \end{bmatrix},
    \quad \hat{\boldsymbol{\pi}}_3 =
    \begin{bmatrix}
    0.145 & 0.364 & 0.492
    \end{bmatrix}, \\[10pt]
    & \hat{\boldsymbol{\pi}}_4 =
    \begin{bmatrix}
    0.222 & 0.577 & 0.201
    \end{bmatrix}
    , \quad \hat{\boldsymbol{\pi}}_5 =
    \begin{bmatrix}
    0.026 & 0.425 & 0.549
    \end{bmatrix} .
  \end{split}
  \nonumber
\end{equation}

The numbers of QDs in the five clusters are as follows.
\begin{table}[h!]
\centering
\begin{tabular}{|c|c|c|c|c|c|}
\hline
 & 1 & 2 & 3 & 4 & 5 \\ \hline
Count & 13 & 51 & 23 & 18 & 23 \\ \hline
\end{tabular}
\end{table}

To analyze the relationship between the 4-cluster and 5-cluster solutions, we present the correspondence of cluster assignments as follows. 

\begin{table}[H]
\centering
\begin{tabular}{rrrrr}
  \hline
 & 1 & 2 & 3 & 4 \\ 
  \hline
1 &  13 &   0 &   0 &   0 \\ 
  2 &   0 &  51 &   0 &   0 \\ 
  3 &   8 &   0 &  15 &   0 \\ 
  4 &   2 &   0 &   0 &  16 \\ 
  5 &   0 &   2 &  12 &   9 \\ 
   \hline
\end{tabular}
\end{table}

Again, it can be seen that Clusters 1 and 2 maintain stability in their composition as the number of clusters increases, whereas the rest of the QDs undergoes progressive subdivisions driven by differences in time spent in medium and high states. Using the 3-cluster model as a reference, the newly formed clusters in the 5-cluster model (Clusters 3--5) predominantly emerge from subdivisions within Cluster 3, with minimal contributions from Clusters 1 and 2.

\end{document}